\documentclass[]{aa}
\usepackage{graphicx}
\usepackage{natbib}
 
\bibpunct{(}{)}{;}{a}{}{,}     
\newcommand{\micron}{$\mu$m}
\newcommand{\ch}[3]{\el{C}_{#1}^{}\el{H}_{#2}^{#3}}
\newcommand{\el}[1]{\mathrm{#1}}

\title{Spatially extended PAHs in circumstellar disks around T Tauri and Herbig Ae stars\thanks{Based on
observations obtained at the European Southern Observatory, Paranal,
Chile, within the observing programs 164.I-0605 (ISAAC May 2002), 
074.C-0413 (NACO, March/April 2005), 075.C-0420 (ISAAC August 2005), 077.C-0668 (VISIR/ISAAC April/May 2006)}}
\titlerunning{Spatially extended PAHs in T Tauri disks}

\author{
V.C. Geers \inst{1}
\and E.F. van Dishoeck\inst{1}
\and R. Visser \inst{1}
\and K.M. Pontoppidan \inst{2}
\and J.-C. Augereau \inst{3}
\and E. Habart \inst{4}
\and A.M. Lagrange \inst{3}
}

\offprints{V.\,C. Geers,\\ \email{vcgeers@strw.leidenuniv.nl}}

\institute{Leiden Observatory, Leiden University, P.O. Box 9513, 2300 RA Leiden, The Netherlands
\and Hubble Fellow, Division of GPS, Mail Code 150-21, California Institute of Technology, Pasadena, CA 91125, USA
\and Laboratoire d'Astrophysique de l'Observatoire de Grenoble, B.P. 53, 38041 Grenoble Cedex 9, France
\and Institut d'Astrophysique Spatiale, 91405 Orsay Cedex, France
}
\authorrunning{Geers et al.}

\date{Received 13 August 2007 / Accepted 11 September 2007}
\abstract
{}
{Our aim is to determine the presence and location of the emission from polycyclic aromatic hydrocarbons (PAHs) towards low and intermediate mass young stars with disks using large aperture telescopes.}
{VLT-VISIR N-band spectra and VLT-ISAAC and VLT-NACO L-band spectra of 29 sources are presented, spectrally resolving the 3.3, 8.6, 11.2, and 12.6 \micron\ PAH features. Spatial-extent profiles of the features and the continuum emission have been derived and used to associate the PAH emission with the disks. The results are discussed in the context of recent PAH-emission disk models.}
{The 3.3, 8.6, and 11.2 \micron\ PAH features are detected toward a small fraction of the T Tauri stars, with typical upper limits between 1 $\times 10^{-15}$ and $5 \times 10^{-17}$ W m$^{-2}$. All 11.2 \micron\ detections from a previous Spitzer survey are confirmed with (tentative) 3.3 \micron\ detections, and both the 8.6 and the 11.2 \micron\ features are detected in all PAH sources. For 6 detections, the spatial extent of the PAH features is confined to scales typically smaller than 0.12--0.34$''$, consistent with the radii of 12-60 AU disks at their distances (typically 150 pc). For 3 additional sources, WL~16, HD~100546, and TY CrA, one or more of the PAH features are more extended than the hot dust continuum of the disk, whereas for Oph IRS~48, the size of the resolved PAH emission is confirmed as smaller than for the large grains. For HD 100546, the 3.3 \micron\ emission is confined to a small radial extent of 12$\pm 3$ AU, most likely associated with the outer rim of the gap in this disk. Gaps with radii out to 10--30 AU may also affect the observed PAH extent for other sources. For both Herbig Ae and T Tauri stars, the small measured extents of the 8.6 and 11.2 \micron\ features are consistent with larger ($\geq$ 100 carbon atoms) PAHs.}
{}


\begin{document}
\maketitle

\section{Introduction}
Mid-infrared (IR) spectroscopy from the ground, with the Infrared Space Observatory (ISO) and recently with the Spitzer Space Telescope has revealed that low (T Tauri) and intermediate mass (Herbig Ae/Be) pre-main-sequence stars often show silicate bands in emission \citep[e.g.,][]{mee01,hon03,boe04,kes06}. Nevertheless only a small fraction of the T Tauri stars, 8\% \citep{gee06}, show clear polycyclic aromatic hydrocarbon (PAH) features, which is low compared to the 54\% detected in Herbig Ae/Be stars \citep{ack04,hab04}. \citet{gee06} argued that their low PAH detection rate is consistent with a 10--100x lower PAH abundance in the disks compared to the ISM.
One difficulty has been to properly identify PAH features in mid-IR spectra in the presence of strong silicate features. The 11.2\,$\mu$m band can be confused with the 11.2\,$\mu$m crystalline forsterite feature and can also be blended with the broad amorphous silicate feature whose strength and spectral width varies with grain size. 

The 3.3\,$\mu$m feature, obtainable through ground-based studies, is another diagnostic of the presence of PAHs in disks. This broad feature is comparatively isolated and is expected to be tightly correlated with the 11.2 \micron\ feature since both involve C--H vibrations, as confirmed by ISO data \cite[e.g.][]{pee04}. Confirming the presence of PAHs in the few T Tauri sources where they have tentatively been seen is the first aim of this paper.

The second aim is to determine whether the PAH emission comes from the disk or rather from an extended remnant envelope of dust around the star. With Spitzer spectroscopy, the spatial resolution of $\sim3''$ at 10 \micron\ in the high-resolution mode corresponds to spatial scales of 300-900 AU for typical nearby star-forming regions ($d$ = 100--300 pc). 
Ground-based observations of the 3.3, as well as the 8.6 and 11.2 \micron\ bands with 8m class telescopes, provide a spatial resolution of $<0.3''$ (at 10 \micron), which is an order of magnitude higher than that of Spitzer.
Recent ground-based, spatially resolved observations of a number of Herbig Ae stars at typical distances of 100--150 pc have shown that the measured PAH features come from regions with sizes typical of a circumstellar disk (radius $<$ 12 AU at 3.3 $\mu$m, $<$100 AU at 11.2 $\mu$m) \citep{hab04,gee05,lag06,dou07}. Spatially resolved, mid-infrared spectroscopy was presented by \citet{boe04} for 3 bright Herbig Ae stars. They found that the 8.6, 11.2, and 12.7 \micron\ features are extended with respect to the continuum emission, on scales of (several) 100 AU. \citet{hab06} presented VLT-NACO observations resolving the 3.3 \micron\ PAH feature above the continuum emission for 4 Herbig Ae/Be sources, with typically 50\% of the intensity coming from within radii smaller than 30 AU. In contrast, observations of WL~16 by \citet{res03} show spatially resolved PAH emission with an extent of 880 $\times$ 400 AU at 1\% of the peak level. 

The above-mentioned observations support the recent modeling results by \citet{hab04}, \citet{gee06}, and \citet{vis07}, which indicate that most ($\sim 80$\%) of the spatially extended PAH emission comes from within a radius $\sim$120--170 AU, with the exception of the 3.3 \micron\ feature for which they predict that half of the emission should originate from $<$50--160 AU from a typical Herbig Ae/Be star, which depends strongly on the size of the PAH molecules. No such data exist yet for T Tauri stars, except for the unusual target IRS~48, a M0 star for which \citet{gee07} measured PAH features with a radial spatial extent of $\sim$ 75--90 AU.

Determining the presence and location of PAHs in the T Tauri disks is significant for several reasons. Due to their high opacity at UV
wavelengths, PAHs can be used as a tracer of the strength of the UV
radiation field, while at the same time their presence in the inner
disk can have a strong influence on the amount of UV that is received
by the outer disk. Also, the strength of PAH features from the outer disk provides a tool
to determine if the disk is flat or flaring \citep{dul04,ack04,hab04}. 
PAHs in the surface layer of the (outer) disk are
an important heating mechanism of the gas through photo-ionization
producing energetic electrons \citep{jon04}. This in turn influences the outer disk chemistry and line emission. PAHs and very small grains (VSGs) also become an important site for H$_2$ formation when the classical grains have grown to \micron\ or mm sizes \citep{jon06}.

High spatial-resolution images taken in narrow band filters centered at particular emission features are one method of constraining the spatial extent of the emitting species, but a good discrimination between the contribution of silicates and PAHs to the mid-infrared excess emission is essential here. Long-slit infrared spectrometers such as ISAAC, VISIR, and NACO, installed on the 8m class VLT telescopes, allow for both spectrally and spatially resolved observations of disks around young (pre-) main sequence stars. 

In this article we present the results of two such surveys, carried out with the VLT-ISAAC and VLT-VISIR instruments. In addition, VLT-NACO L-band spectroscopy is obtained on two sources, making use of adaptive optics to obtain even higher spatial resolution. The aim of this work is to obtain a limited survey of the 3.3 and 11.2 \micron\ features, with a focus on low-mass T Tauri stars with disks, to study the usefulness of the 3.3 \micron\ feature as a PAH tracer compared with the 11.2 \micron\ feature and to study the spatial extent of the PAH emission in the context of the disk.

\section{Observations and data reduction}
\subsection{Source selection}
The sample contains 19 T Tauri stars and 10 Herbig Ae/Be stars and was selected as follows: 17 T Tauri stars in the nearby star-forming regions Chamaeleon, Lupus, Ophiuchus, and Serpens were chosen from the sample observed with Spitzer Infrared Spectrograph (IRS) in the context of the ``Cores to Disks'' (c2d) Legacy program \citep{eva03}. All their sources with definite and tentative PAH detections \citep{gee06} were chosen, including 4 Herbig Ae stars. In addition, as part of a backup program during pointing limited nights with strong winds, 4 additional young stars in the Serpens and Corona Australis star-forming regions, found in IRAC + MIPS imaging surveys, were observed with ISAAC. One source, HD~100546, was observed in an earlier ISAAC program. Three sources, IRS~48, WL~16, and EC~82, were added to the VISIR N-band program after serendipitous discovery of the 3.3 \micron\ PAH band in the L-band spectra in an ISAAC survey of embedded sources \citep{pon03}. The final source list is given in Table\,\ref{tab:obssum}. 

\subsection{ISAAC L-band spectroscopy}
L-band spectroscopy was obtained with ISAAC, the Infrared Spectrometer And Array Camera, installed at the VLT Antu at ESO's Paranal Observatory in Chile, in the nights of June 16, 2000 (HD~100546),  August 9--14, 2005, and April 17--18, 2006 in the low-resolution ($\lambda / \Delta \lambda$ = 600) spectroscopic mode in the spectral domain 2.8--4.2 $\mu$m using a $0.6\arcsec\times120\arcsec$ slit. The telescope was operated using a chop throw of 20\arcsec\ and a nod throw of 20\arcsec. Because of chopping, these observations are only sensitive to spatially extended emission of at most 10\arcsec. For flux calibration and telluric line correction, several standard stars were observed.
The data were reduced using our own IDL routines, first described in \citet{pon03}. The individual frames were corrected for the nonlinearity of the detector array and distortion corrected using a startrace map; bad pixels and cosmic ray hits were removed before co-adding, using a shift-and-add procedure to correct for telescope jitter. From the combined frames, both the positive and the two negative spectral traces were extracted and co-added. For correction of the telluric features, the extracted source spectrum was divided by the similarly extracted, standard star spectrum using an optimal small wavelength shift and an exponential airmass correction between the source and the standard, thus requiring that the pixel-to-pixel noise on the continuum of the final science spectrum is minimized. Before this exponential airmass correction was applied, the detector and filter response curves were removed to obtain a spectrum of the true atmospheric absorption. These curves were obtained by fitting an envelope to both the standard star spectrum and a spectrum of the approximate atmospheric transmission and by taking the ratio of the two envelopes.
Flux calibration was performed by dividing the science spectrum by an observed standard star and multiplying by a blackbody with the effective temperature of the standard star. Airmass correction was applied by multiplying the flux of the science target by the factor $10^{-0.4 \mathrm{Ext}_{\mathrm{L}} (\mathrm{AM}_{\mathrm{st}} - \mathrm{AM}_{\mathrm{sc}})}$
where we use for the L-band atmospheric extinction, \mbox{$\mathrm{Ext}_{\mathrm{L}} = 0.08$} magn. $\times$ airmass$^{-1}$, the value provided by ESO on the ISAAC webpage\footnote{http://www.eso.org/instruments/isaac/\\imaging\_standards.html}. 
The flux calibration is estimated to be accurate to 30\%. The spectrum is wavelength-calibrated relative to the atmospheric transmission spectrum and is accurate to $\sim$0.003 \micron. 
Note that the observation campaign in August 2005 suffered from poor weather conditions, with strong winds and variable seeing. Part of the nights were pointing limited to the North due to strong winds, which led to a relatively large fraction of the backup program sample of Serpens sources being observed. A summary of the presented observations is given in Table\,\ref{tab:obssum}. 

\subsection{NACO L-band spectroscopy}
L-band spectroscopy was obtained with NAOS-CONICA  installed at the VLT-Yepun at ESO's Paranal Observatory in Chile, in service mode, in the nights of 25--26 March and 10 April of 2005 in the medium-resolution ($R=700$) spectroscopic mode in the spectral domain 3.20--3.76 \micron\ using a 0.172$''$x28$''$ slit. The data were reduced using our own IDL routines, following a similar procedure as for the ISAAC data. The spectrum was wavelength-calibrated relative to the atmospheric transmission spectrum and is accurate to $\sim$0.008 \micron. Flux calibration was performed using ISAAC spectro-photometry, estimated to be accurate to 30\%. One source, WL~16, was observed with the long-slit of the spectrograph aligned in 2 settings, both perpendicular and parallel to the semi-major axis of the disk, which is at a position angle of $60\pm2\degr$ \citep{res03}. A summary of the observations is given in Table\,\ref{tab:obssum}.

\subsection{VISIR N-band spectroscopy}
N-band spectroscopy was obtained with VISIR installed at the VLT-Melipal at ESO's Paranal Observatory in Chile, in the nights of 3--7 May of 2006 in the low-resolution ($R \sim$ 350) spectroscopic mode in the spectral domain 7.7--12.5 $\mu$m using a $0.75\arcsec\times32.3\arcsec$ slit. All sources were observed with the 8.5 and 11.4 \micron\ settings to cover the main 8.6 and 11.2 \micron\ PAH features. In addition, a few sources were observed with the 8.1 and/or 12.2 \micron\ settings to cover the red wing of the 7.7 \micron, as well as the 12.7 \micron\ PAH feature. Most of the nights were characterized by strong winds, pointing limited to the south and highly variable seeing. The telescope was operated using chop-throws and nod-throws of 8\arcsec. Because of chopping, these observations are only sensitive to spatially extended emission of at most 4\arcsec. Several standard stars were observed for flux calibration and telluric-line correction.
The data were reduced using our own IDL routines, following a similar procedure as for the ISAAC data. Flux calibration was performed using Spitzer IRAC 8 \micron\ photometry where possible. The uncertainty in the flux calibration was estimated to be 30\%. The spectrum was wavelength-calibrated using the ESO VISIR pipeline (v.\ 1.3.7) by comparison with atmospheric lines and is accurate to $\sim 10^{-4}$ \micron.
A summary of the observations is given in Table\,\ref{tab:obssum}.
\begin{table*}
\centering
\tiny
\caption{Summary of observations.}
\label{tab:obssum}
\begin{tabular}{lllllll}
\hline
\hline
Target          & RA (2000)          & Dec (2000)        & Date        & Sp. Type & Dist. (pc) & Ref.\\
\hline							   	     
{\bf ISAAC L-band} \\					   		     
\hline						       
SX~Cha          & 10 55 59.73 & $-$77 24 39.9 & 17-04-2006  & M0       & 178   & K06; W97   \\   
SY~Cha          & 10 56 30.45 & $-$77 11 39.3 & 18-04-2006  & M0.5     & 178   & K06; W97   \\  
WX~Cha          & 11 09 58.74 & $-$77 37 08.9 & 19-04-2006  & K7-M0    & 178   & K06; W97   \\ 
HD~98922        & 11 22 31.67  & $-$53 22 11.4   & 19-04-2006  & B9       & $>$540& A98; H78   \\ 
HD~101412       & 11 39 44.46  & $-$60 10 27.9   & 19-04-2006  & B9.5     & 160   & T94; A05   \\ 
HD~100546       & 11 33 25.44 & $-$70 11 41.2 & 16-06-2000  & B9       & 103   & A98; A98   \\ 
T~Cha        & 11 57 13.49 & $-$79 21 31.4 & 18-04-2006  & G8       & 66    & A93; A98   \\ 
IRAS~12535-7623 & 12 57 11.73 & $-$76 40 11.1 & 19-04-2006  & M0       & \ldots& K06; -  \\ 
HT~Lup          & 15 45 12.86 & $-$34 17 30.6 & 09-08-2005  & K2       & 140   & H94; H93   \\ 
GW~Lup          & 15 46 44.73 & $-$34 30 35.5 & 14-08-2005  & M2-M4    & 140   & H94; H93   \\ 
SZ~73           & 15 47 56.94 & $-$35 14 34.7 & 13-08-2005  & M0    & 140     & H94; H93  \\ 
GQ~Lup          & 15 49 12.10 & $-$35 39 05.1 & 10-08-2005  & K7-M0    & 140     & H94; H93 \\ 
HD~141569       & 15 49 57.75 & $-$03 55 16.4 & 13-08-2005  & B9.5-A0  & 100   & AU04; AU04 \\ 
IM~Lup          & 15 56 09.22 & $-$37 56 05.8 & 14-08-2005  & M0       & 150--360     & H94; H93,K01 \\ 
RU~Lup          & 15 56 42.30 & $-$37 49 15.4 & 10-08-2005  & K7-M0    & 150--360     & H94; H93,K01 \\ 
DoAr~24E        & 16 26 23.26 & $-$24 20 59.8 & 15-08-2005  & K0       & 125     & K06; G89 \\ 
Em* SR~21A          & 16 27 10.28 & $-$24 19 12.7 & 15-08-2005  & G2.5     & 125   & P03; G89 \\ 
Em* SR~9            & 16 27 40.29 & $-$24 22 04.0 & 10-08-2005  & K5-M2    & 125   & L99; G89\\ 
Haro~1-17       & 16 32 21.93 & $-$24 42 14.8 & 14-08-2005  & M2.5     & 125   & A93; G89   \\ 
V1121~Oph       & 16 49 15.30 & $-$14 22 08.7 & 10-08-2005  & K5 & -     & V00; - \\ 
Wa~Oph~6        & 16 48 45.62 & $-$14 16 36.0 & 11-08-2005  & K & -     & G07; - \\ 
VV~Ser          & 18 28 47.86 & +00 08 39.8 & 11-08-2005  & A0V      & 259   & M01; S96   \\ 
CoKu~Ser~G6     & 18 29 01.23 & +00 29 33.0 & 11-08-2005  & K3     & 259     &  C79; S96 \\ 
HD~176386       & 19 01 38.89 & $-$36 53 27.0 & 14-08-2005  & B9.5 & 140    & G93; S00 \\ 
TY~CrA          & 19 01 40.79 & $-$36 52 34.2 & 14-08-2005  & B9	     & 140$^{a}$     &  V00; S00 \\ 
T~CrA           & 19 01 58.78 & $-$36 57 49.9 & 10-08-2005  & F0e     & 140$^{a}$     & F84; S00 \\ 
\hline
\bf{VISIR N-band} \\
\hline
SY~Cha          & 10 56 30.45 & $-$77 11 39.3 & 06-05-2006  & M0       & 178   & K06; W97   \\ 
WX~Cha          & 11 09 58.74 & $-$77 37 08.9 & 06-05-2006 & K7-M0    & 178   & K06; W97 \\ 
HD~98922        & 11 22 31.67  & $-$53 22 11.4   & 07-05-2006  & B9       & $>$540& A98; H78\\ 
HD~101412       & 11 39 44.46  & $-$60 10 27.9   & 07-05-2006  & B9.5     & 160   & T94; A05\\ 
T~Cha           & 11 57 13.49 & $-$79 21 31.4 & 04-05-2006 & G8       & 66    & A93; A98  \\ 
WL~16           & 16 27 02.34  & $-$24 37 27.2   & 06-05-2006 & B8-A7        & 125   & L99; G89  \\ 
SR~21A          & 16 27 10.28  & $-$24 19 12.7   & 06-05-2006 & G2.5     & 125   & P03; G89  \\ 
Oph~IRS~48      & 16 27 37.19 & $-$24 30 35.0 & 12-06-2005  & M0       & 125   & O07; G89   \\ 
VV~Ser          & 18 28 47.86 & +00 08 39.8 & 07-05-2006  & A0V      & 259   & M01; S96   \\ 
EC~82           & 18 29 56.80 & +01 14 46.0 & 07-05-2006   & M0       & 259     & K06; S96     \\ 
\hline
\bf{NACO L-band} \\
\hline
WL~16 set 1       & 16 27 02.34  & $-$24 37 27.2   & 25-03-2005 & B8-A7        & 125   & L99; G89   \\ 
WL~16 set 2       & 16 27 02.34  & $-$24 37 27.2   & 26-03-2005 & B8-A7        & 125   & L99; G89  \\ 
WL~16 set 3       & 16 27 02.34  & $-$24 37 27.2   & 10-04-2005 & B8-A7        & 125   & L99; G89  \\ 
Oph~IRS~48      & 16 27 37.19 & $-$24 30 35.0 & 10-04-2005  & M0       & 125   & O07; G89  \\ 
\hline
\end{tabular}
\begin{list}{}{}
\item $^{a}$: all CrA sources assumed to be at same distance as that derived by S00 for HD~176386.
\item References for spectral type, distance:
A93: \citet{alc93},
A98: \citet{anc98},
A04: \citet{ack04},
AU04: \citet{aug04},
A05: \citet{ack05},
C79: \citet{coh79},
C06: \citet{com07},
D97: \citet{dun97},
F84: \citet{fin84},
G89: \citet{geu89},
G93: \citet{gra93},
G07: \citet{gra07},
H78: \citet{hou78},
H93: \citet{hug93},
H94: \citet{hug94},
K01: \citet{knu01},
K06: \citet{kes06},
L99: \citet{luh99},
M01: \citet{mor01},
O07: Oliveira et al.\ (2007, in prep.),
P03: \citet{pra03},
R03: \citet{res03},
S00: \citet{sie00},
S96: \citet{str96},
T94: \citet{the94},
V00: \citet{val00},
W97: \citet{whi97}, 
\end{list}
\end{table*}

\subsection{Measuring spatial extent}
\label{ssec:measspatext}
The FWHM of the spatial profile of the ISAAC spectra was derived from the co-added 2D spectral images, by fitting a Gaussian profile for each wavelength bin. For the ISAAC data, distortion correction was not performed because this correction was found to introduce a semi-sinusoidal pattern. The spatial extent of the PAH feature at 3.3 \micron\ was measured with respect to the extent of the disk continuum emission at 3.3 \micron, interpolated between the continua points at 3.1 and 3.5 \micron. The extent in arcseconds was converted to a radial extent from the center in AU for sources with known distances. 

Because of the variable seeing during a large part of our program, the atmospheric seeing was assumed to dominate the FWHM of the ISAAC and VISIR observations and the spatial blurring introduced by instrumental optics ignored. The standard star was assumed to be an unresolved point source and, where possible, its measured FWHM was used as a measure of the point-spread-function. 
Only for the case of HD~100546 were the conditions good enough to match the source and standard spectra by applying differential airmass and seeing corrections.

For the VISIR and NACO data, the extent of the spatial profile was derived as the standard deviation of the flux distribution $F_i$ over the spatial pixels $x_i$. For each wavelength bin, the standard deviation $\sigma^2$ was calculated as $\Sigma (x_i - C)^2 \times Fi/\Sigma F_i$, with centroid $C = \Sigma (x_i \times F_i)/\Sigma F_i$. Assuming a normal distribution, 3 $\sigma$ corresponded to 99\% of the spatial extent of the flux. 
The NAOS-CONICA observations of WL~16 and IRS~48 were taken using the adaptive optics (AO) system. This resulted in high angular-resolution observations, which were typically close to the diffraction limit ($\sim 0.1''$ at 3.3 \micron). 

\section{Results and discussion}
\subsection{PAH detections and statistics}
\begin{figure}
  \centering
  \includegraphics[width=\columnwidth]{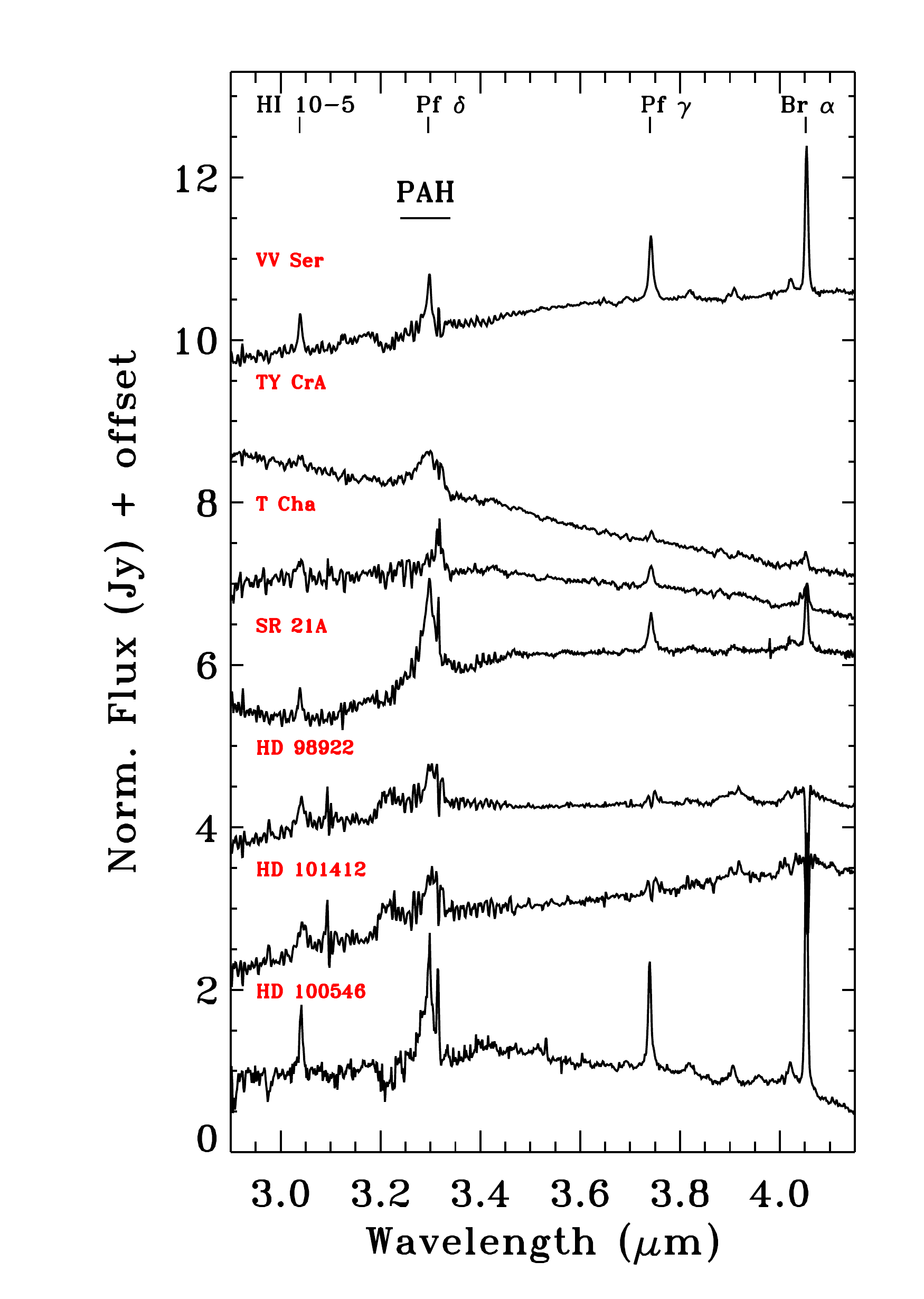}
  \caption{ISAAC L-band spectra of sample with 3.3 \micron\ PAH detection.}
  \label{fig:isaacspec}
\end{figure}

\begin{figure}
  \centering
  \includegraphics[width=\columnwidth]{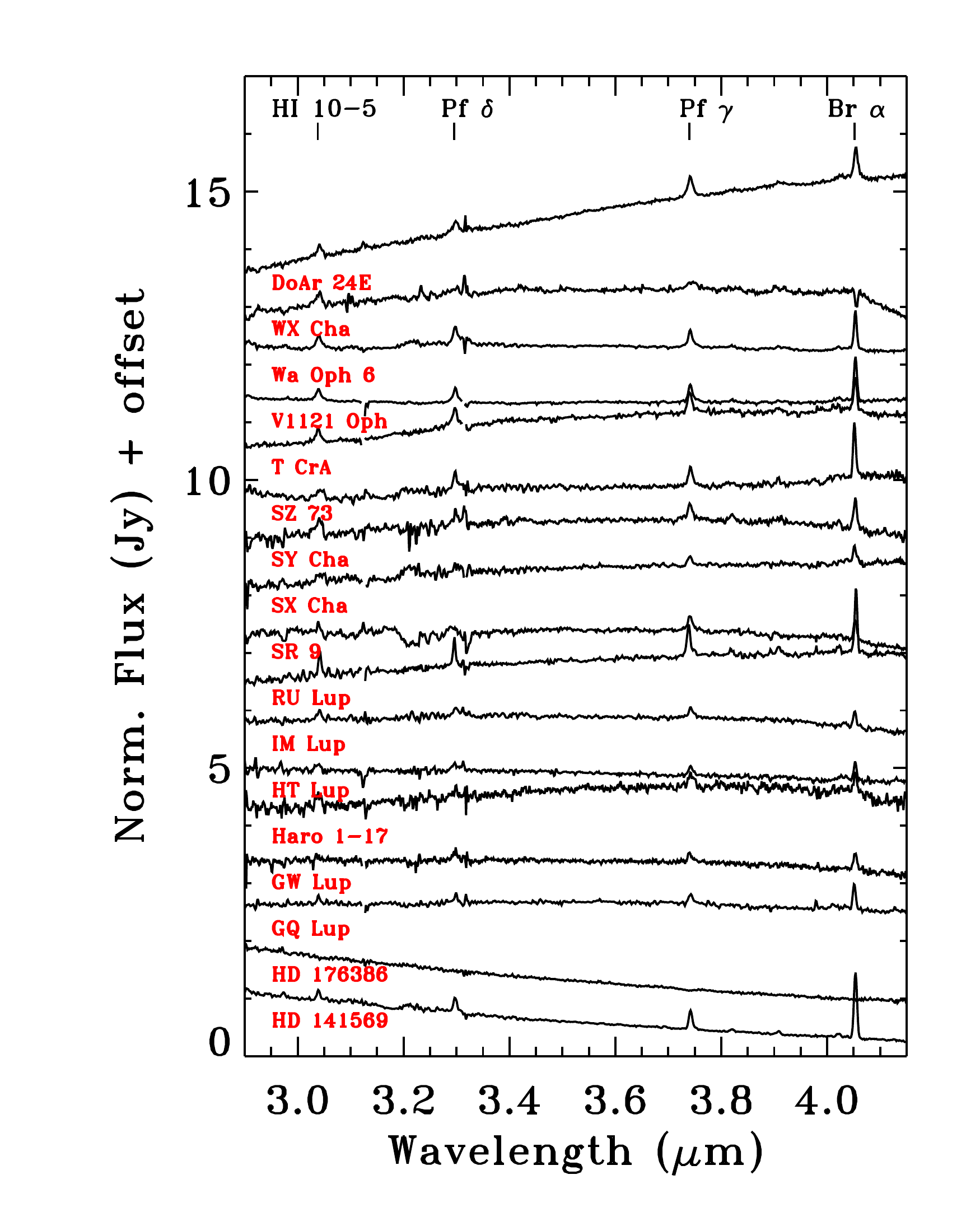}
  \caption{ISAAC L-band spectra of sample without 3.3 \micron\ PAH detection.}
  \label{fig:isaacspec_nopah}
\end{figure}

\begin{figure}
  \centering
  \includegraphics[width=\columnwidth]{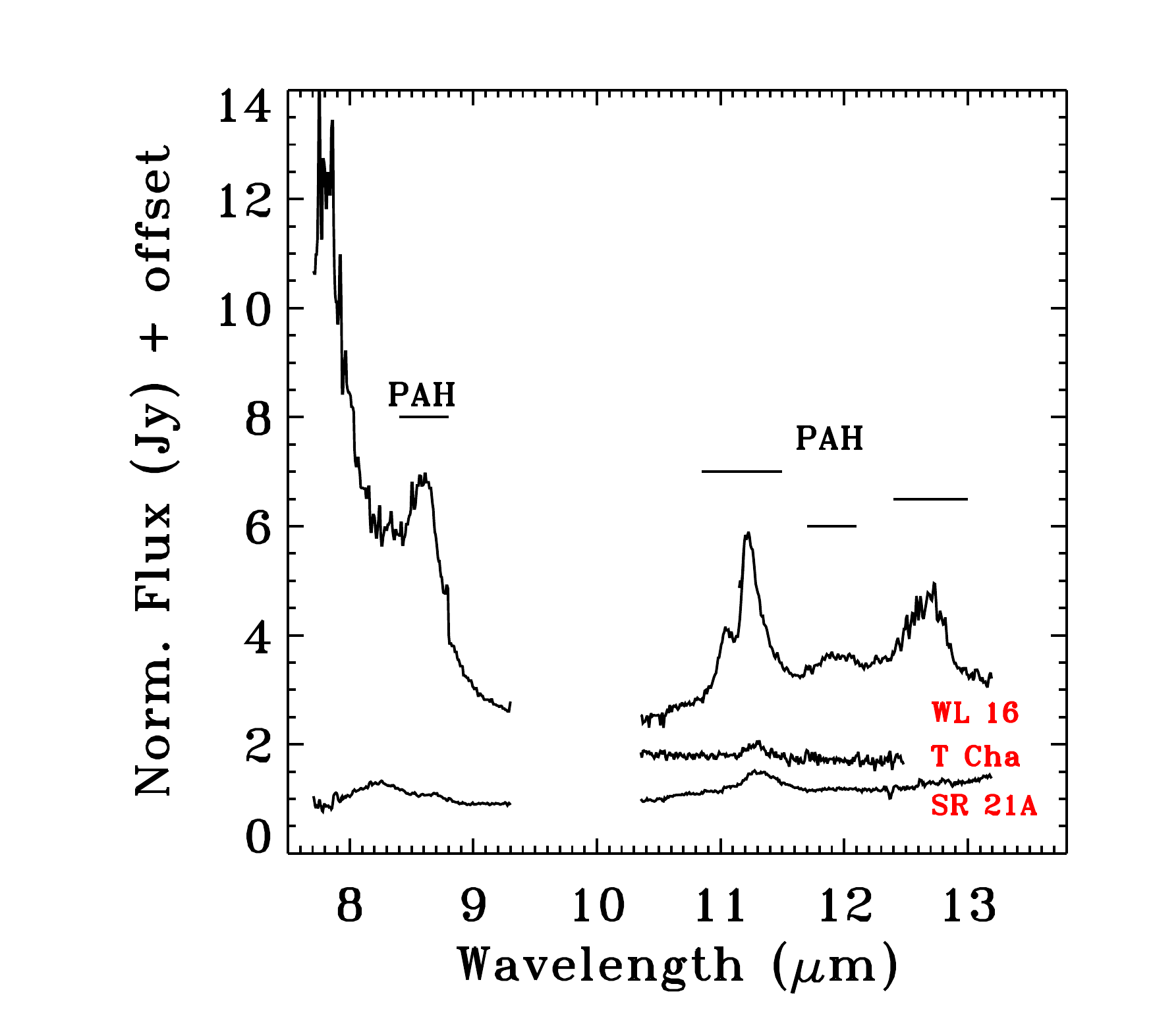}
  \caption{VISIR N-band spectra of the sources with PAH detections.}
  \label{fig:visirdet2}
\end{figure}

\begin{figure}
  \centering
  \includegraphics[width=\columnwidth]{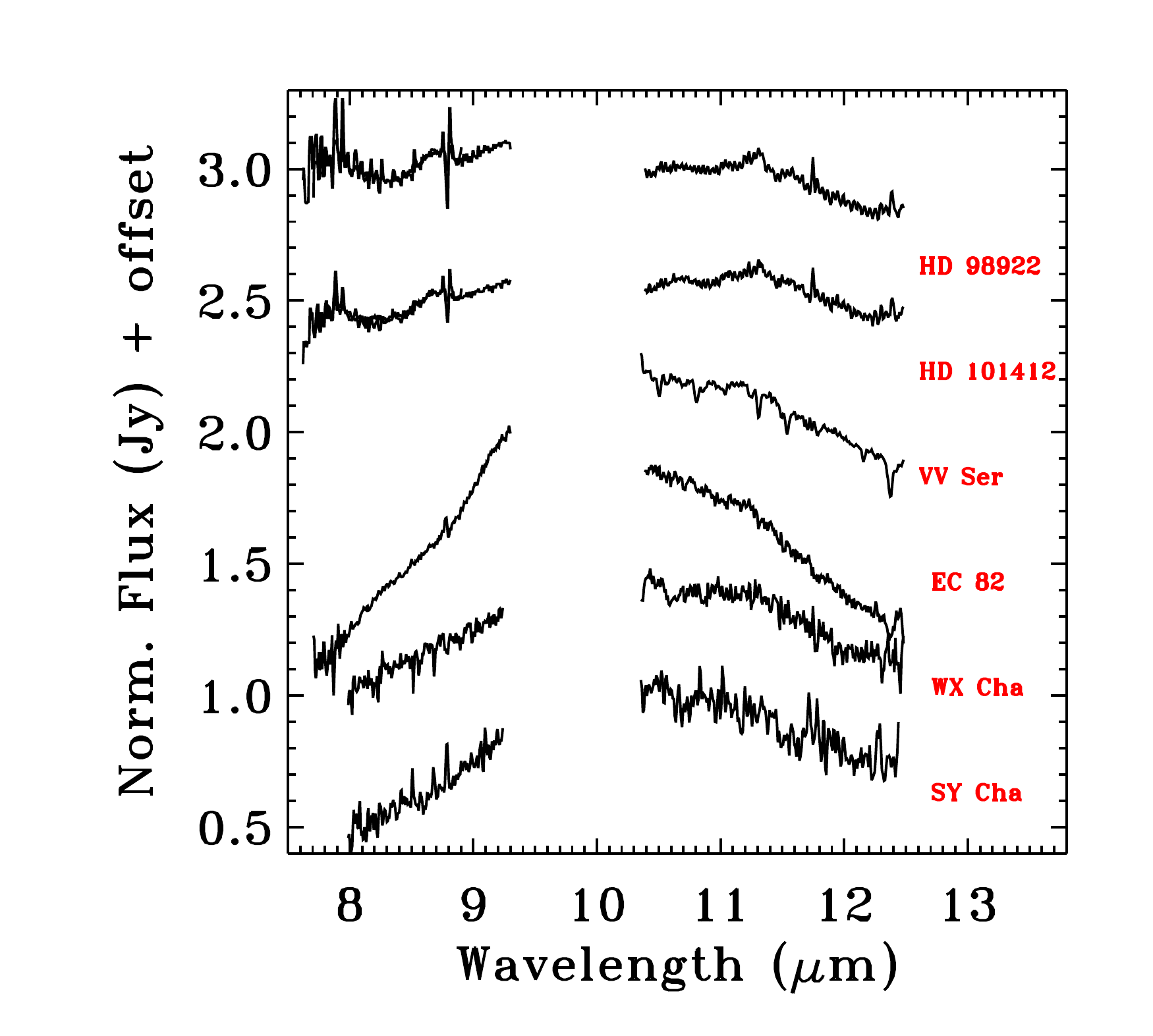}
  \caption{VISIR N-band spectra of the sources with marginal (top 4) and no PAH detections (bottom 2).}
  \label{fig:visirdet}
\end{figure}

The L-band spectra are presented in Figs.\ \ref{fig:isaacspec} and \ref{fig:isaacspec_nopah}. Table \ref{tab:featflux} lists the line intensities of the detected PAH features.
The N-band spectra are shown in Figs.\ \ref{fig:visirdet2} and \ref{fig:visirdet}. 

We detect the 3.3 \micron\ PAH feature in 6 out of 24 sources, of which 4 Herbig Ae/Be stars (TY~CrA, HD~100546, HD~98922, HD~101412) and 2 T Tauri stars (SR~21A, IRS~48). For VV~Ser and T~Cha, a marginal feature is found. The HI Br~$\alpha$, Pf~$\gamma$, Pf~$\delta$, and 10-5 transition lines are seen in the majority of the sample, of which the HI Pf~$\delta$ line lies superposed on the 3.3 \micron\ feature. 
Note that we do not detect the 3.3 \micron\ feature toward the Herbig Ae star HD~141569. A tentative 11.2 \micron\ feature was presented by \citet{syl96} and later clearly detected by \citet{slo05}. This absence of the 3.3 and presence of the 11.2 \micron\ feature is consistent with the model prediction of \citet{li03}, where the 3.3 \micron\ feature is weak due to the high degree of ionization in this disk \citep{all99}.

In the N-band spectra, PAH features are detected in 3 Herbig Ae/Be (WL~16, HD~98922, HD~101412) and 4 T Tauri stars (T~Cha, SR~21A, IRS~48, EC~82), shown in Figs.~\ref{fig:visirdet2} and \ref{fig:visirdet}. In all cases, except EC~82, both the 8.6 and the 11.2 \micron\ features are detected. For WL~16 and IRS~48, additional features at 11.9 and 12.8 \micron\ are clearly seen. 

The 3.3 \micron\ PAH band is observed in only a very small number of T Tauri stars, 2 out of 18. Compared to \citet{gee06}, all Herbig Ae and T Tauri stars with 11.2 \micron\ Spitzer detections also show 3.3 \micron\ features in this L-band survey, confirming the detections. In addition, we find 3.3 \micron\ PAH features in a few sources that were not included the Spitzer sample: IRS~48, WL~16, and HD~100546, all of which were known to have PAHs based on previous data \citep{mal98,res03,gee05}. Conversely, we have not found any new T Tauri disks with PAHs at 3.3 \micron. The sensitivity varied significantly throughout the nights, but typical upper limits between 1 $\times 10^{-15}$ and $5 \times 10^{-17}$ W m$^{-2}$ are obtained for the 3.3~\micron\ PAH feature, comparable to the Spitzer study of \citet{gee06} for the 11.2 \micron\ feature.
The 11.2 \micron\ feature is equally observed in only a small fraction of our T Tauri sample observed with VISIR. The Spitzer detections of \citet{gee06} are confirmed, and derived line intensities are consistent for most sources.  The VISIR spectrum of VV~Ser suffers from poor telluric correction and does not show a clear 11.2 \micron\ feature, in contrast to the Spitzer data. The spectra of HD~98922 and HD~101412 confirm the same weak, broad, and confused 11.2 \micron\ line noted to be possibly due to crystalline silicates \citep{gee06}. For both sources, \citet{kes06} confirmed the presence of crystalline silicates from detection of the 23 and 33 \micron\ crystalline silicate features. However, the  8.6 \micron\ detections in both VISIR spectra strengthens the conclusion that PAHs also contribute to the 11.2 \micron\ feature. 
In five sources, both the 3.3 and 11.2 \micron\ PAH features are detected (HD~98922, HD~101412, IRS~48, WL~16, SR~21A). One source, T~Cha, has 11.2 \micron\ clearly detected but 3.3 \micron\ only marginally. 

\subsection{Spatial extent}
\label{ssec:spatextresult}
\begin{figure}
  \centering
  \includegraphics[width=\columnwidth]{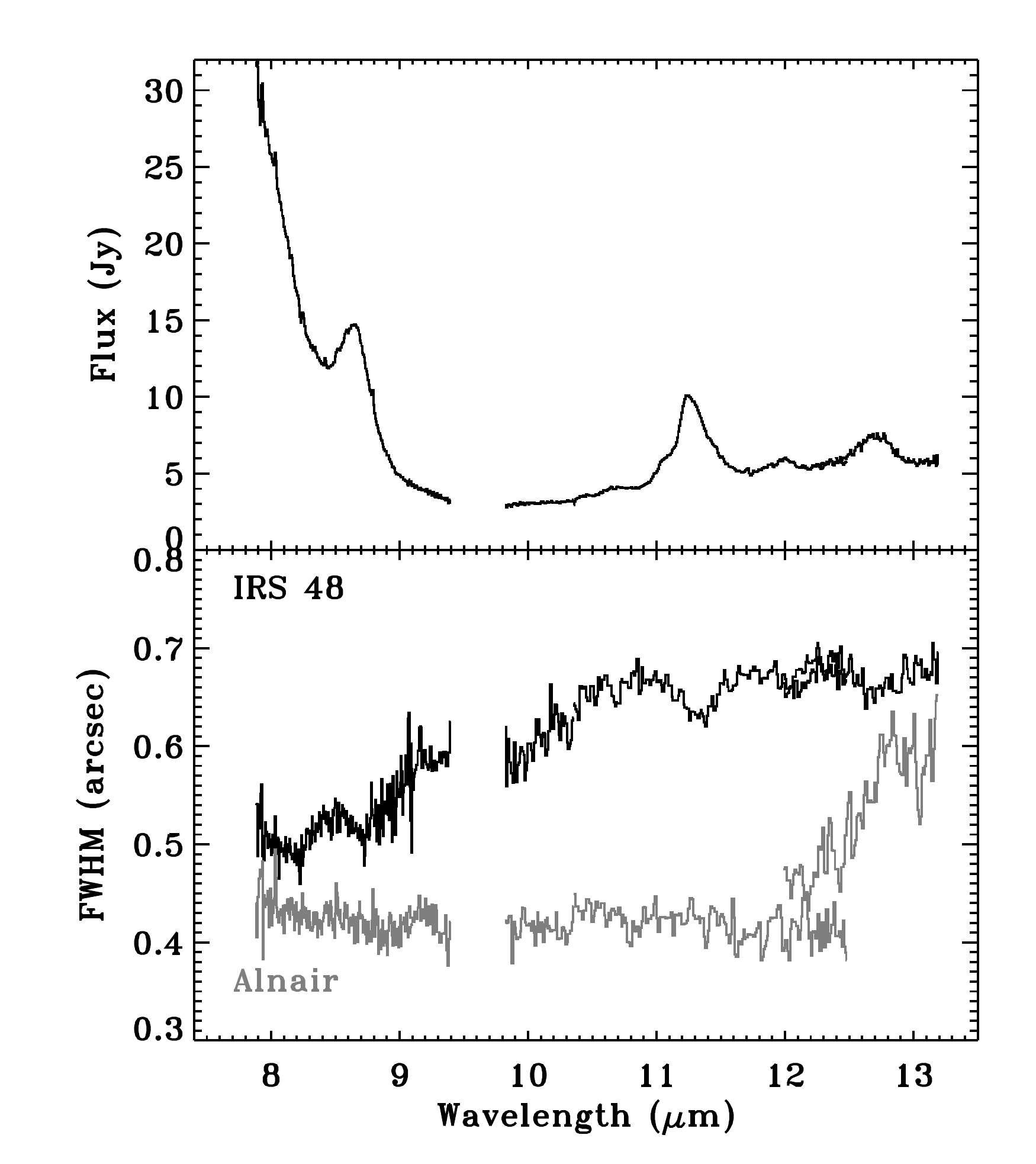}
  \caption{({\bf Upper panel}) VISIR N-band spectrum of IRS~48. ({\bf Lower panel}) FWHM of spatial profile of IRS~48 and standard star Alnair.}
  \label{fig:irs48}
\end{figure}

Table \ref{tab:featextent} lists the radial spatial extent of the PAH features in half-width at half-maximum (HWHM). Spatial extent profiles are shown for a few sources, HD~100546, TY~CrA, SR~21A, WL~16, and IRS~48, in the lower panels of Figs.~\ref{fig:hd100}--\ref{fig:irs48naco1}. These figures show the spectrum in the top panel and in the bottom panel the diameter (full-width at half-maximum, or FWHM) of the spatial profile fit as a function of wavelength, for both the science source and the telluric and PSF standard star. 

The majority of sources with PAH detections show no spatial extent of the features beyond the surrounding continuum emission from the disk at 3.3 \micron, confining the source to the same spatial extent as the disk continuum emission. In many cases the upper limit on the spatial extent is close to that of SR~21A, with a radius of $0.33''$ (41 AU at $d=125$ pc), an example of a source with very strong unresolved PAH features (Fig.~\ref{fig:sr21}). These observations constrain the origin of the PAH emission from T Tauri sources to the circumstellar disk. 

The hydrogen emission lines, 10-5, Pf $\delta$, Pf $\gamma$, and Br $\alpha$, are detected in the majority of the spectra and are spatially unresolved in all cases. These lines are expected to originate from hot ionized gas in the very inner parts of the star+disk system, close to the surface of the central star.

The maximum spatial extent can be observed either when the disk is fully face-on or when it is inclined at a moderate degree with the slit of the spectrograph aligned with the semi-major axis of the disk. For most sources, the inclination and position angle of the disk are unknown and thus the derived spatial extents are in most cases treated as a lower limit for the disk component. For a typical $45\degr$ inclination, assuming a circular disk and aligning perpendicular to the apparent semi-major axis, this would lead to an underestimate of the extent by a factor of 1.4. In the following, individual cases are discussed.

\begin{table*}
\centering
\scriptsize
\caption{Summary of PAH feature intensity in W m$^{-2}$ and presence of HI lines.}
\label{tab:featflux}
\begin{tabular}{llllll}
\hline
\hline
Source 		&  H lines & 3.3 \micron\                         & 8.6 \micron\ & 11.2 \micron\ & 12.7 \micron\ \\
\hline            
SX~Cha          &  Y	   & $\leq 1.9 \times 10^{-15}$   & - 	 & -	& -   \\
SY~Cha          &  Y	   & $\leq 8.0 \times 10^{-17}$   & -	 & -	& -   \\
WX~Cha          &  Y	   & $\leq 2.7 \times 10^{-16}$   & N    & N	& -   \\
HD~98922        &  Y	   & $4.4 \times 10^{-14}$	  & Y	 & Y	& -   \\
HD~101412       &  Y	   & $2.8 \times 10^{-15}$ $^{a}$ & Y	 & Y	& -   \\
HD~100546       &  Y	   & Y  	                  & -    & -	& -   \\
T~Cha           &  Y	   & $\leq 2.5 \times 10^{-15}$	  & -	 & -    & -   \\
IRAS~12535-7623 &  Y	   & N  	                  & -	 & -	& -   \\
HT~Lup	        &  Y	   & $\leq 1.3 \times 10^{-15}$   & -	 & -	& -   \\
GW~Lup          &  Y	   & $\leq 2.4 \times 10^{-16}$   & -	 & -	& -   \\
SZ~73           &  Y	   & $9.4 \times 10^{-16}$ $^{a}$ & -	 & -	& -   \\
GQ~Lup          &  Y	   & $\leq 5.4 \times 10^{-16}$   & -	 & -	& -   \\
HD~141569       &  Y	   & $\leq 6.9 \times 10^{-16}$   & -	 & -	& -   \\
IM~Lup          &  Y	   & $\leq 1.9 \times 10^{-15}$   & -	 & -	& -   \\
RU~Lup          &  Y	   & $\leq 5.7 \times 10^{-15}$ $^{a}$ & -	 & -	& -   \\
DoAr~24E        &  Y	   & $\leq 4.1 \times 10^{-16}$   & -	 & -	& -   \\
WL~16           &  Y	   & $\leq 9.6 \times 10^{-17}$   & Y	 & Y    & Y   \\
Em* SR~21A          &  Y	   & $3.5 \times 10^{-15}$	  & Y	 & Y	& T   \\
Oph~IRS~48       &  Y	   & $2.4 \times 10^{-15}$	  & Y	 & Y	& Y   \\
Em* SR~9            &  Y	   & $\leq 9.7 \times 10^{-16}$ $^{a}$ & -	 & - & -   \\
Haro~1-17       &  Y	   & $\leq 4.5 \times 10^{-17}$   & -	 & -	& -   \\
V1121~Oph       &  Y	   & $1.2 \times 10^{-15}$ $^{a}$ & -	 & -	& -   \\
Wa~Oph~6        &  Y	   & $2.3 \times 10^{-15}$ $^{a}$ & -	 & -	& -   \\
VV~Ser          &  Y	   & $\leq 9.8 \times 10^{-16}$   & -	 & T	& -   \\
CoKu~Ser~G6     &  T	   & N  	                  & -	 & -	& -   \\
EC~82           &  -	   & -  	                  & N	 & T	& -   \\
HD~176386       &  N	   & $\leq 4.2 \times 10^{-16}$   & -	 & -	& -   \\
TY~CrA          &  Y	   & $4.5 \times 10^{-15}$	  & -	 & -    & -   \\
T~CrA           &  Y	   & $\leq 3.9 \times 10^{-15}$   & -	 & -	& -   \\
\hline		   
\end{tabular} 
\begin{list}{}{}
\item - : not observed
\item $^{a}$ Line intensity affected by HI Pf $\delta$ line
\end{list}
\end{table*}  
\begin{table*}
\centering
\scriptsize
\caption{Summary of the radial spatial extent of the PAH feature (half-width at half-maximum).}
\label{tab:featextent}
\begin{tabular}{lllllllll}
\hline
\hline
Source 		&   \multicolumn{2}{c}{3.3 \micron} & \multicolumn{2}{c}{8.6 \micron} & \multicolumn{2}{c}{11.2 \micron} & \multicolumn{2}{c}{12.7 \micron}  \\
\hline            
\multicolumn{9}{l}{PAH detected and spatially resolved} \\
\hline
HD~100546  & $0.12''$ & 12 AU & - & -   & - & -   & - & -\\
TY~CrA          & $0.39''$ & 55 AU & -    & -    & - & -    & -    & -\\
WL~16           & - & - & $0.43''$ & 54 AU & 0.49$''$ & 61 AU & 0.46$''$ & 58 AU \\
IRS~48          & $0.11''$ & 14 AU & $0.25''$ & $31$ AU   & $0.32''$ & $40$ AU  & $0.33''$& $41$ AU \\
\hline
\multicolumn{9}{l}{PAH detected but spatially unresolved} \\
\hline
HD~98922        & $\leq 0.24''$ & $\leq$ 130 AU$^{a}$ & $\leq 0.17''$ & $\leq 92$ AU & $\leq 0.15''$ & $\leq 81$ AU  & - & - \\
HD~101412       & $\leq 0.24''$  & $\leq$ 38 AU & $\leq 0.16''$ & $\leq$ 26 AU & $\leq 0.15''$ & $\leq$ 24 AU & - & -\\
Em* SR~21A          & $\leq 0.33''$ & $\leq$ 41 AU &$\leq 0.16''$ & $\leq$ 20 AU &$\leq 0.17''$&  $\leq$ 21 AU & $\leq 0.15''$ & $\leq$ 19 AU\\
T~Cha           & $\leq 0.34''$ & $\leq$ 22 AU & -    & -  & $\leq 0.19''$ & $\leq$ 13 AU & - & - \\
EC~82           & - & - & $\leq 0.18''$    & $\leq$ 47 AU & $\leq 0.23''$    & $\leq$ 60 AU & - & -\\
\hline		   
\end{tabular}
\begin{list}{}{}
\item $^{a}$ : uncertain distance to source
\end{list}
\end{table*}  
	     
\subsubsection{HD~100546}
\begin{figure}
  \centering
  \includegraphics[width=\columnwidth]{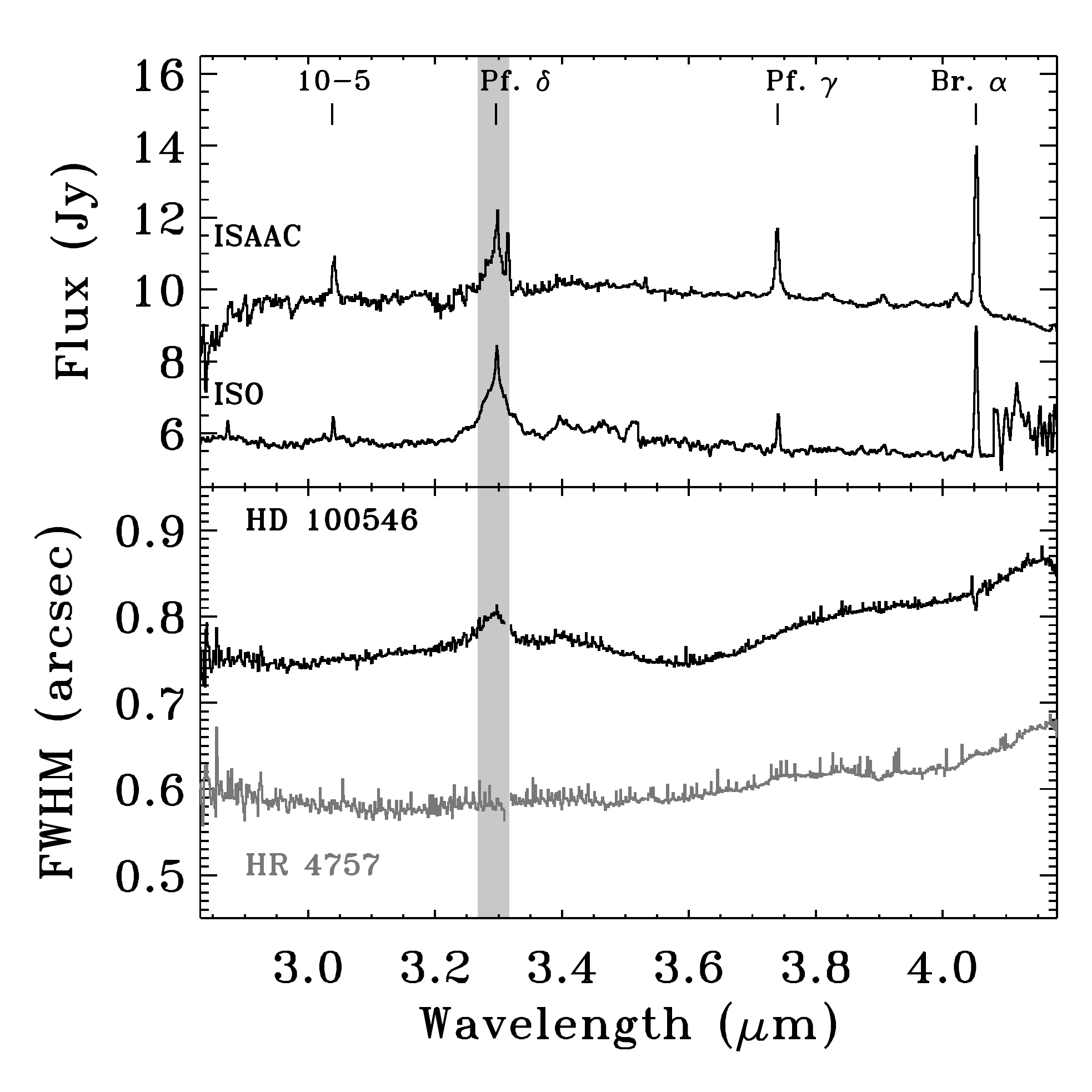}
  \caption{({\bf Upper panel}) ISAAC L-band spectrum of HD~100546, scaled by a factor 1.25 to match the ISO spectrum and then shifted by +4\,Jy for clarity. The grey band highlights the position and FWHM of the 3.3 \micron\ PAH feature as determined in this paper. ({\bf Lower panel}) FWHM of the spatial profile, as extracted from 2D spectral image, of HD100546 ({\it black line}) and its corresponding standard HR~4757 ({\it grey line}). The FWHM of the latter agrees with that of the HD~100546 continuum after correction for airmass and seeing differences.}
\label{fig:hd100}
\end{figure}
In HD~100546 (Fig. \ref{fig:hd100}) the long-slit of ISAAC was aligned to the semi-major axis of the disk, to measure the maximum spatial extent of the emission. The 3.3 \micron\ PAH feature is spatially resolved, with a radial extent of $0.40''$, found to be slightly larger than the neighboring 3 \micron\ continuum extent.
After correcting for the difference in airmass and seeing with respect to the standard star spectrum, we find that the feature has a spatial extent of 0.12$'' \pm 0.032''$ and that the continuum emission is essentially unresolved. Assuming a distance of 103 pc, this corresponds to a radial extent of $12 \pm 3$ AU. 

HD~100546 is one of the most nearby isolated Herbig Be stars for which a circumstellar disk has been firmly established through space and ground-based spectroscopy and imaging. Physical parameters for HD~100546 and its distance have been derived from Hipparcos observations \citep{anc98}: it is a B9Ve star with a mass of 2.4 M$_\odot$. The circumstellar disk has been resolved through imaging at visible \citep{gra01}, near-infrared \citep{pan00,aug01a}, mid-infrared \citep{liu03}, and millimeter \citep{wil03} wavelengths. 
\citet{aug01a} observed an elliptical structure, consistent with an extended inclined disk, based on HST/NICMOS2 coronagraphic imaging at 1.6 \micron\ and derived values for the disk inclination and position angle adopted here.

\citet{bou03} modeled the spatial distribution and chemical composition of the dust around HD~100546 and find that the data required the presence of a small grain component at $\sim$200\,K, inconsistent with a uniform flaring disk \citep{dul01}. Instead, they propose that the disk of HD~100546 has a gap between $\sim$ 1 and 10 AU and is `puffed up' at a radius of about 10 AU. 
Based on these results, they conclude that the observations of the spectral energy distribution (SED) of  HD~100546 are consistent with a largely cleared-out inner region between the inner rim and 10 AU and that most of the disk material is located in the outer parts of the disk. This conclusion was subsequently confirmed by \citet{ack06b}. Our derived spatial extent of the 3.3 \micron\ feature of $12\pm3$ AU would place the PAHs responsible for the 3.3 \micron\ emission at the puffed-up rim at the outer edge of the cleared out region where a relatively large fraction of UV radiation is intercepted.

\citet{hab06} report the presence of a 5-10 AU radial gap in the 3.3 \micron\ PAH spatial distribution, and a clear extension to at least 50 AU radially, which is larger than the typical extent (30 AU) found in their sample. This gap is not seen in our ISAAC data because of our lower spatial resolution. Our ISAAC HWHM of the spatial extent of $\sim$ 12 AU is smaller than their full 50 AU extent, presumably from our lower sensitivity to weaker extended emission with ISAAC. Their PAH distribution is consistent with our finding that the PAH emission is dominated by the dust rim at the outer edge of the gap. 

\subsubsection{TY~CrA}
\begin{figure}
  \centering
  \includegraphics[width=\columnwidth]{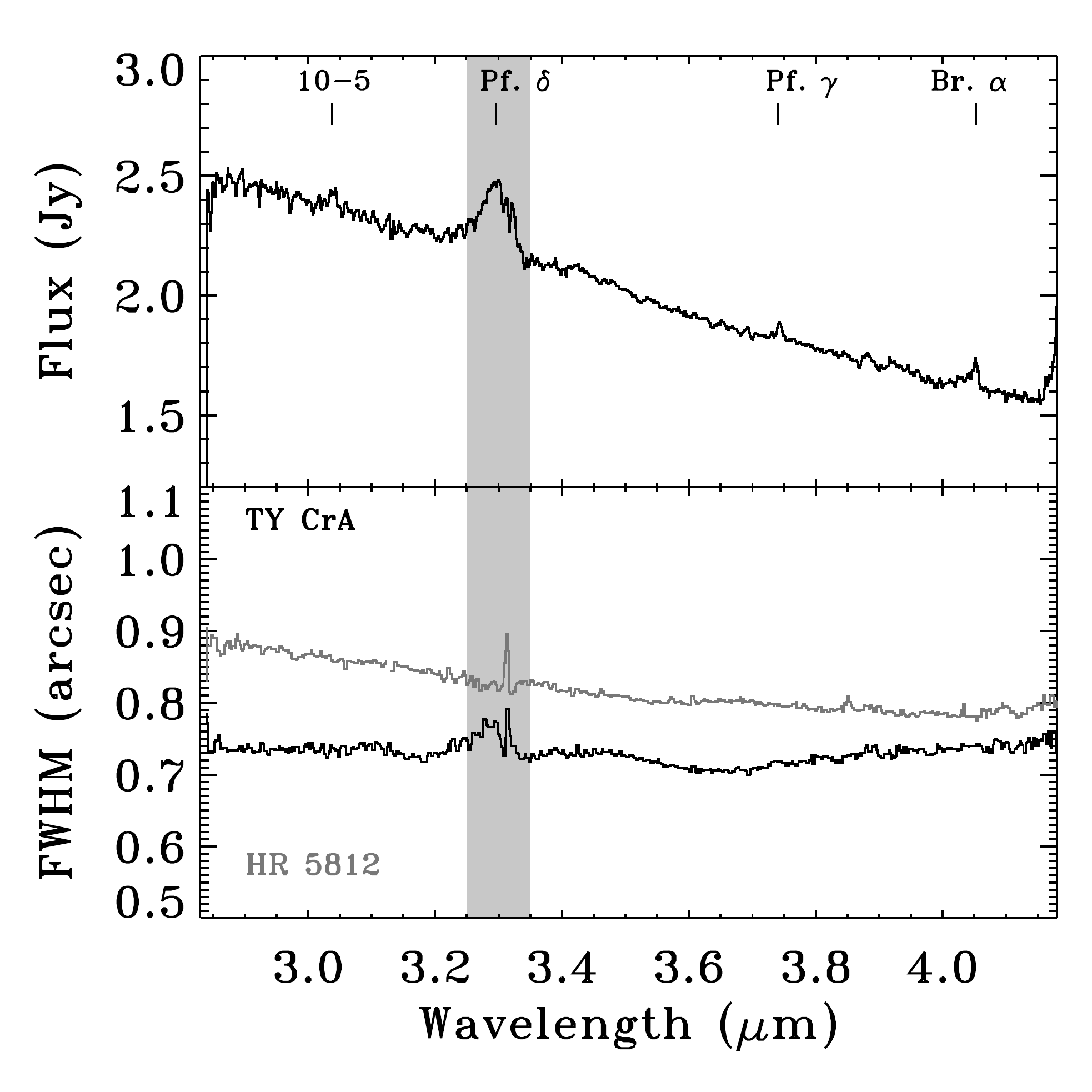}
  \caption{({\bf Upper panel}) ISAAC L-band spectrum of TY~CrA. ({\bf Lower panel}) FWHM of spatial profile of TY~CrA and standard star HR~5812.}
  \label{fig:tycra}
\end{figure}
\begin{figure}
  \centering
  \includegraphics[width=\columnwidth]{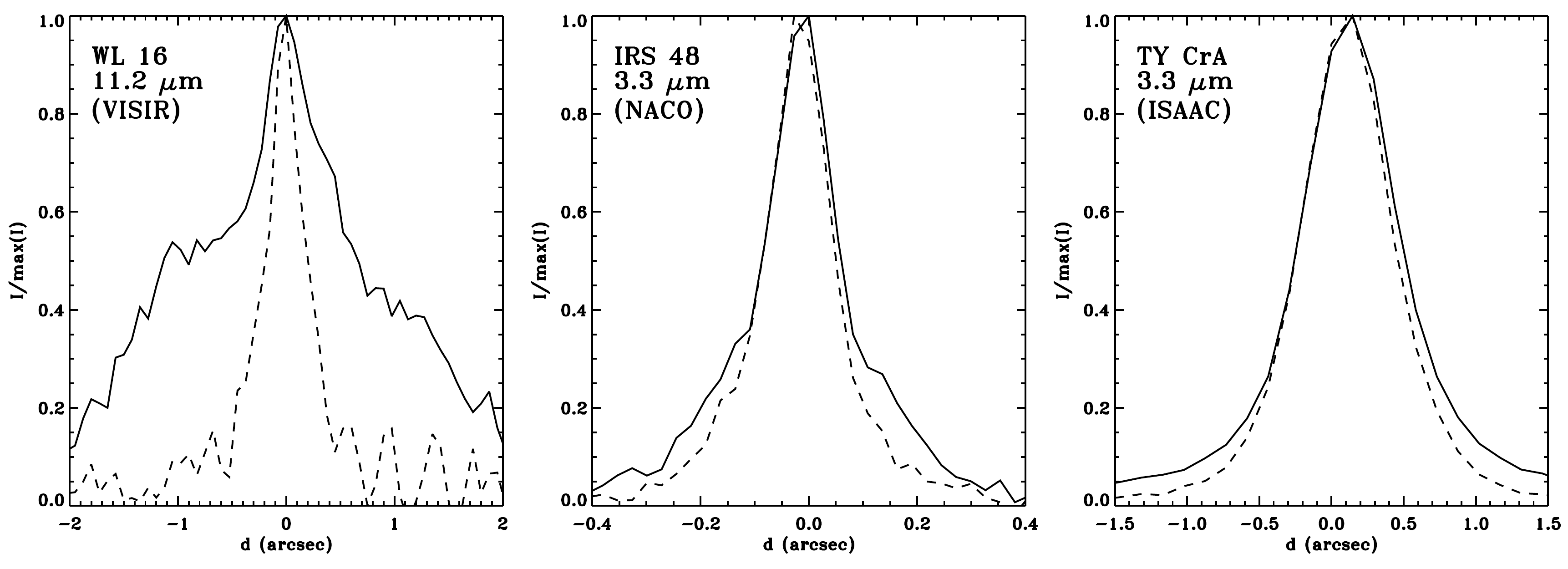}
  \caption{Observed spatial emission profiles for WL~16 (left), IRS~48 (middle), and TY CrA (right), for the PAH feature at indicated wavelength (solid line) and continuum emission at 10.51, 3.56, and 3.35 \micron\ respectively (dashed line).}
  \label{fig:spatprof}
\end{figure}
For TY~CrA (Fig. \ref{fig:tycra}), the 3.3 \micron\ feature is found to be spatially resolved (0.78$''$, corresponding to a radial extent of 54 AU) compared with the underlying 3.3 \micron\ disk continuum (0.73$''$, or 51 AU radially). The coordinates of this compact component match within 1$''$ with the optical multiple star system and with the nearest bright 2MASS Ks-band source.  TY~CrA is a very strong X-ray source, surrounded by a strong reflection nebula. Unpublished Spitzer/IRAC observations show a large extended nebulosity at 8 \micron\ around this source (Allen, priv.\ comm.), confirming that PAHs are present on even larger scales as shown by \citet{sie00} at 11.28 \micron. The spatial emission profile of the 3.3 \micron\ feature is shown in Fig.\ \ref{fig:spatprof}, which confirms the presence of the weak extended PAH emission. PAH features have been previously observed toward this star with ISO \citep{kla06,ack06}. NACO imaging revealed this system to be a possible quadruple system of low-mass M type stars \citep{cha03}, possibly affecting the circumstellar disk(s) structure and resulting PAH emission. 

\subsubsection{WL~16}
\begin{figure}
  \centering
  \includegraphics[width=\columnwidth]{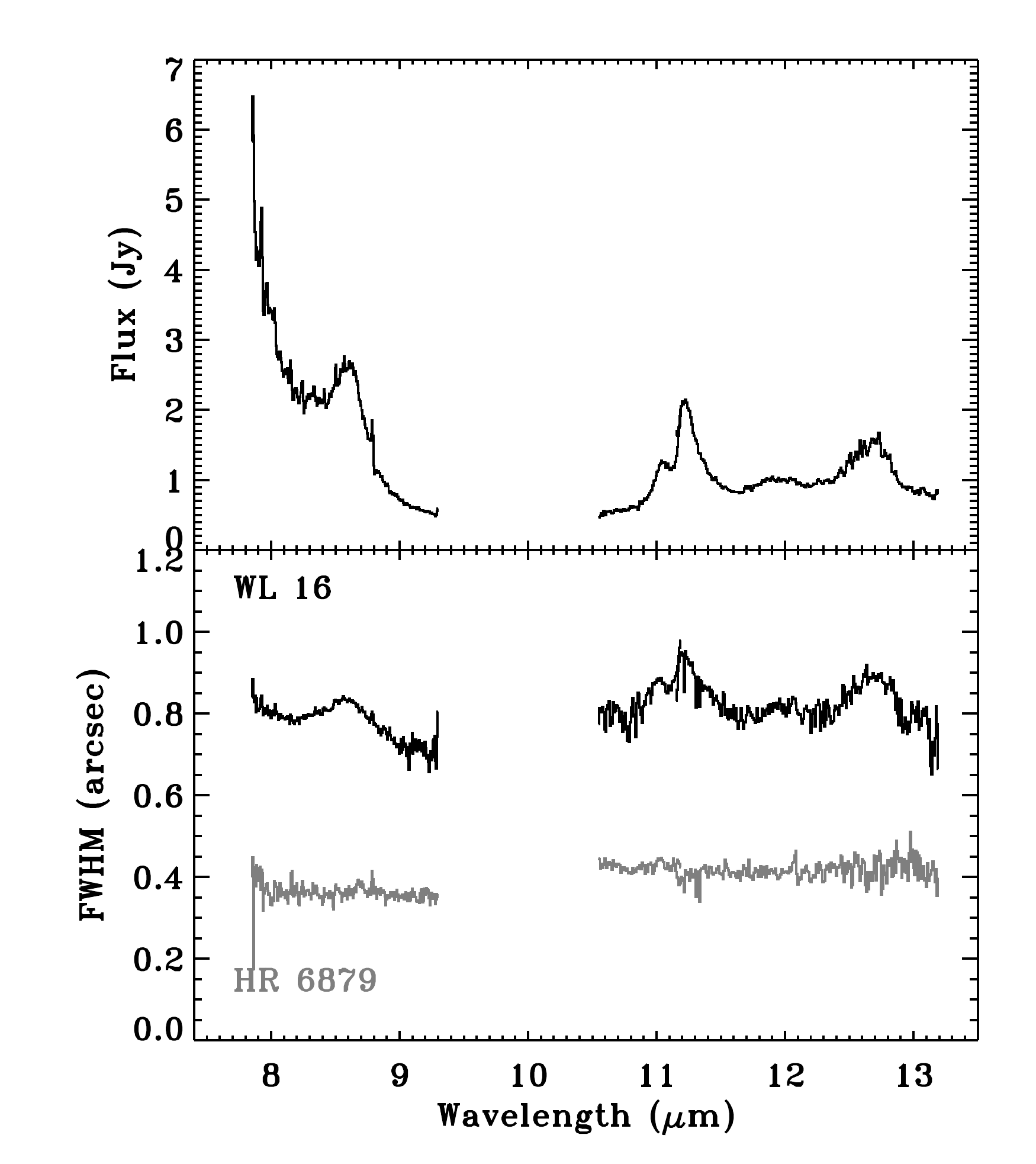}
  \caption{({\bf Upper panel}) VISIR N-band spectrum of WL~16. ({\bf Lower panel}) FWHM of spatial profile of WL~16 and standard star HR~6879.}
  \label{fig:wl16}
\end{figure}
\begin{figure}
  \centering
  \includegraphics[width=\columnwidth]{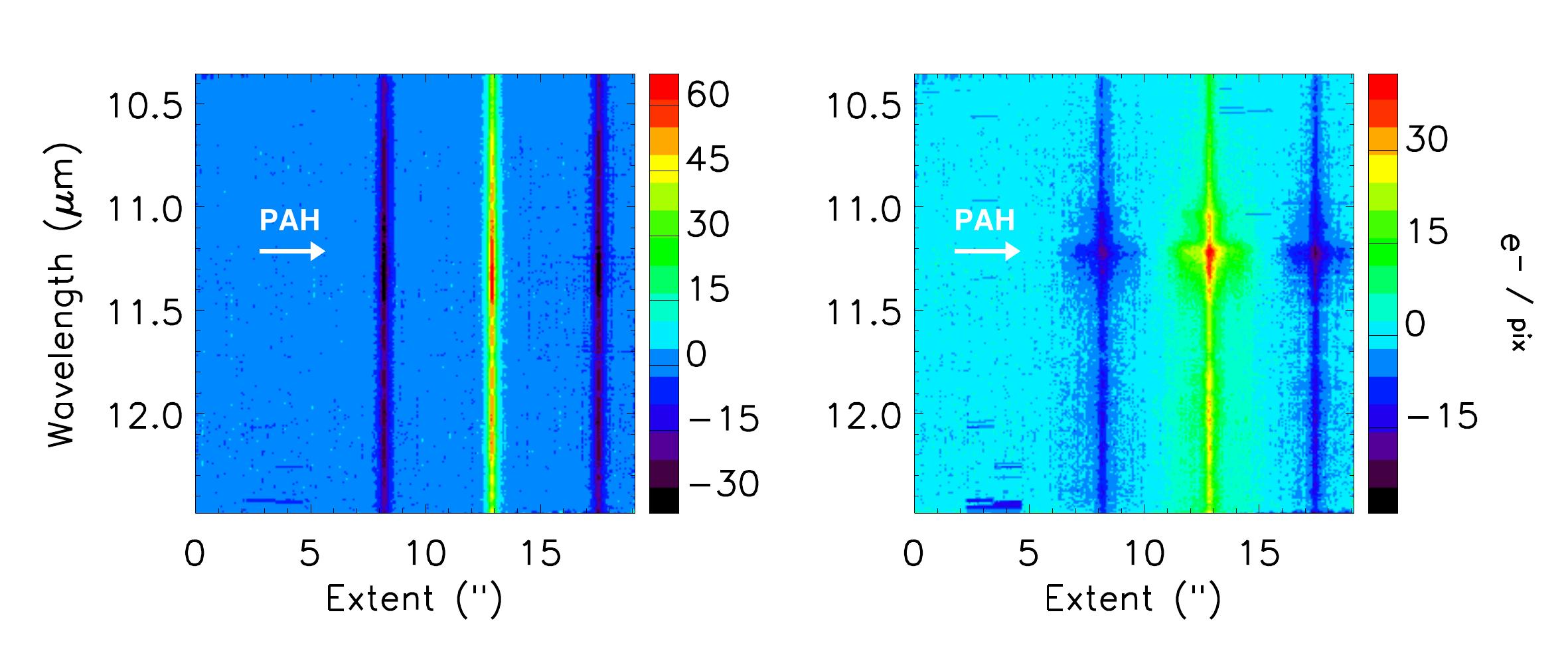}
  \caption{VISIR 2D spectral image of SR~21A (left) and WL~16 (right); 1 pixel = 0.02719$''$.}
  \label{fig:wl16_2dspec}
\end{figure}
\begin{figure}
  \centering
  \includegraphics[width=\columnwidth]{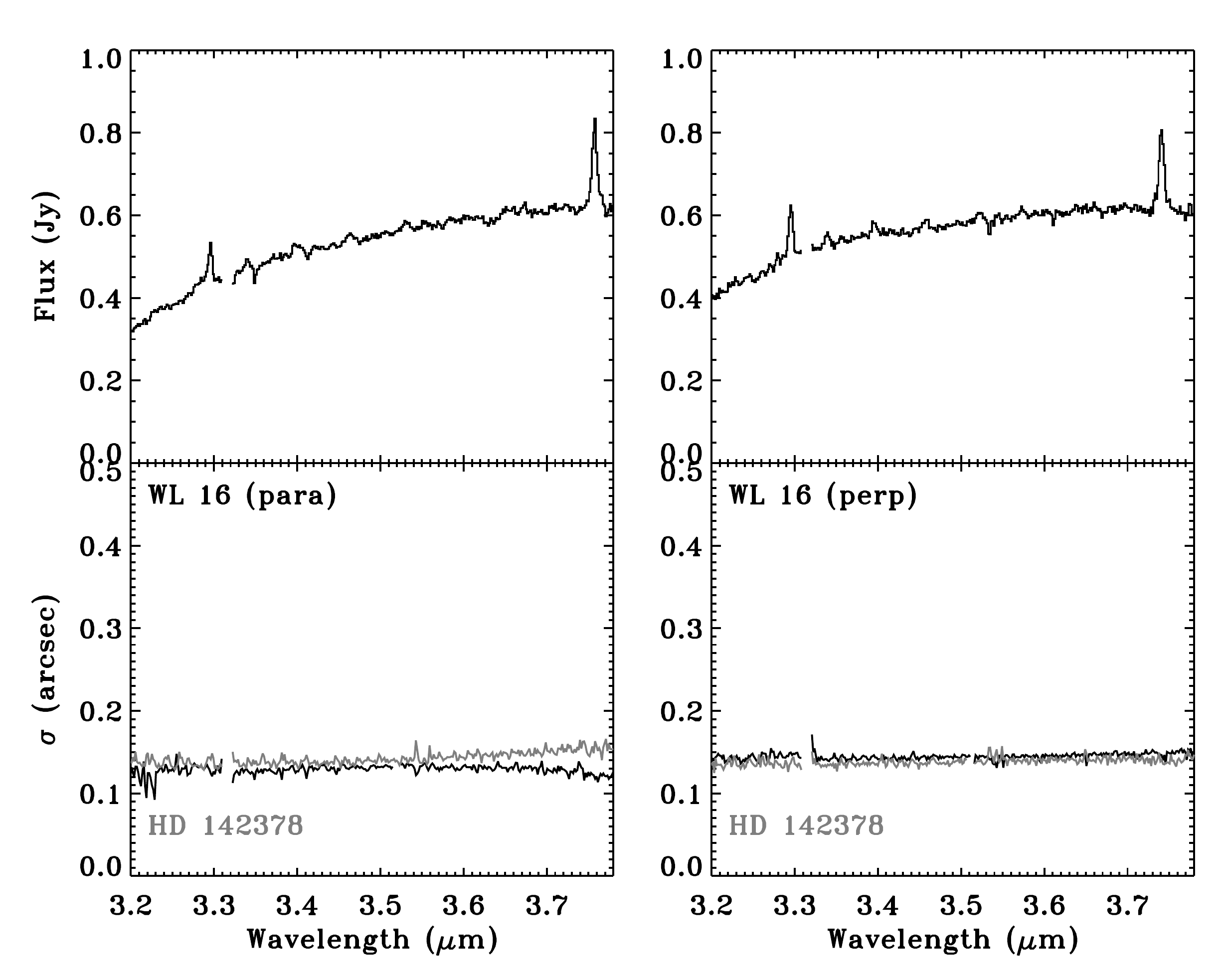}
  \caption{({\bf Upper-left panel}) NACO L-band spectrum of WL~16, slit parallel to semi-major axis of the disk. ({\bf Lower-left panel}) FWHM of spatial profile of the ``parallel'' observation of WL~16 (black) and standard star HD~142378 (grey). ({\bf Upper-right panel}) NACO L-band spectrum of WL~16, slit perpendicular to semi-major axis of the disk. ({\bf Lower panel}) FWHM of spatial profile of the ``perpendicular'' observation of WL~16 (black) and standard star HD~142378 (grey).}
  \label{fig:wl16naco1}
\end{figure}
For WL16, the disk continuum emission at 8--13 \micron\ is resolved with a radial extent of about 0.4$''$ (50 AU) (Figs.~\ref{fig:wl16} and \ref{fig:wl16_2dspec}). The 8.6, 11.2, and 12.7 \micron\ PAH features are resolved with respect to the continuum, with radial spatial extents of 0.43$''$ (54 AU), 0.49$''$ (61 AU), and 0.46$''$ (58 AU). The spatial emission profile is shown in the left panel of Fig.\ \ref{fig:spatprof}. The PAH emission at 10--13 \micron\ has broad wings in the spatial direction, extending beyond the spatial extent derived using the statistical derivation described in Sec.\ \ref{ssec:measspatext}, illustrating how in certain cases it is hard to quantify the spatial extent in a single number. The extent varies with wavelength, which is not expected from uniform diffuse background PAH emission.
Deriving the extent alternatively as 1\% of the peak level, we find a value of 80--140 AU, which is about a factor 2.5 lower than that found by \citet{res03} of 440 by 220 AU radially, based on Keck imaging covering 7.9 to 24.5 \micron. Given the reported semi-major over semi-minor axis ratio of 0.466, the effect of non-alignment of the VISIR slit with the semi-major axis can be at most a factor of 2.

NACO L-band spectroscopy was taken in two settings, aligning the long-slit parallel to the semi-major axis and perpendicular. The 3.3 \micron\ feature is undetected in both orientations (Fig.\ \ref{fig:wl16naco1}). The radial spatial extent of the continuum at 3.3 \micron\ is small and very similar between both orientations. 
The non-detection at 3.3 \micron\ would be consistent with predominantly ionized PAHs \citep{all99,li01}. 

\subsubsection{SR~21A}
\begin{figure}
  \centering
  \includegraphics[width=\columnwidth]{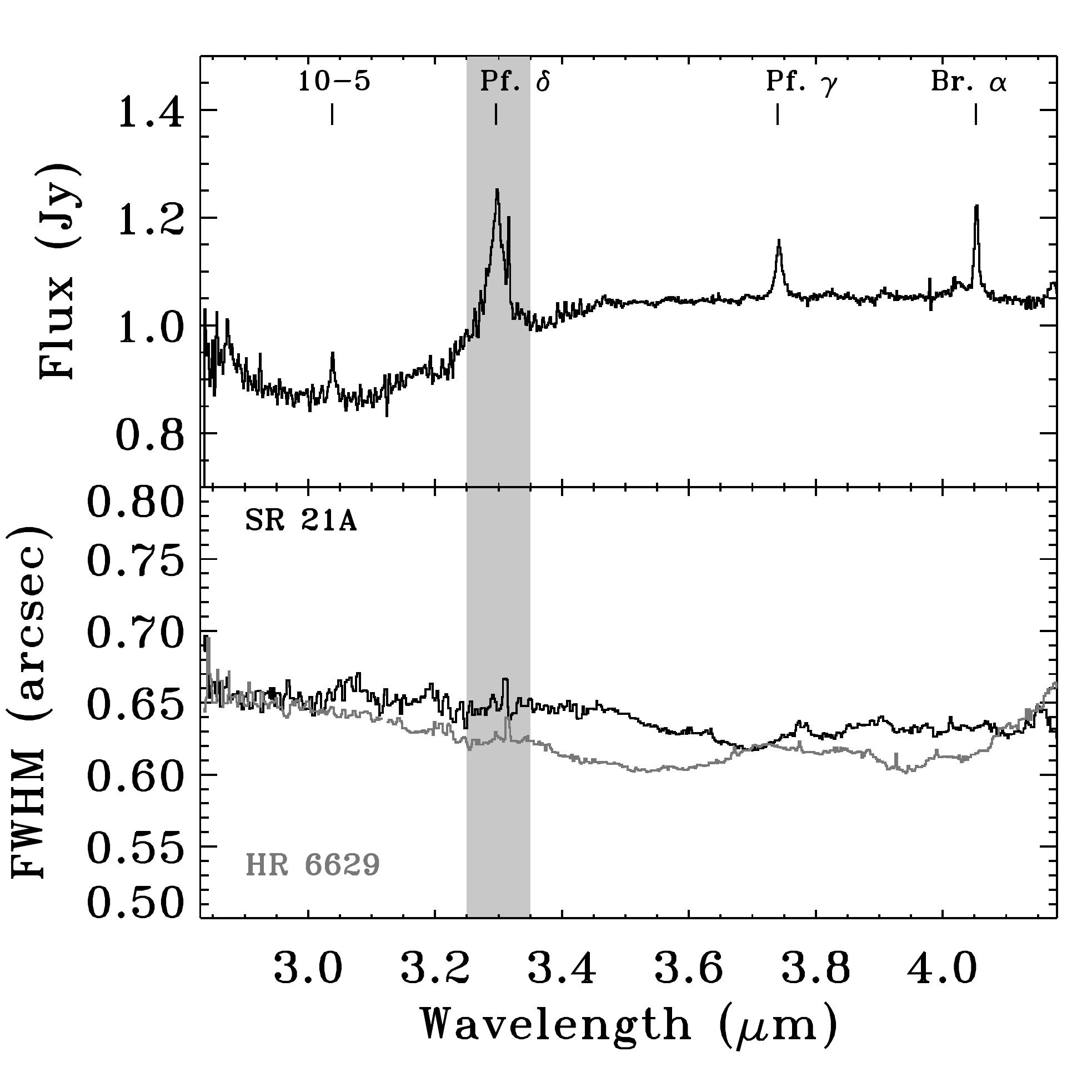}
   \caption{({\bf Upper panel}) ISAAC L-band spectrum of SR~21A. ({\bf Lower panel}) FWHM of spatial profile of SR~21A and standard star HR~6629.}
  \label{fig:sr21}
\end{figure}
\begin{figure}
  \centering
  \includegraphics[width=\columnwidth]{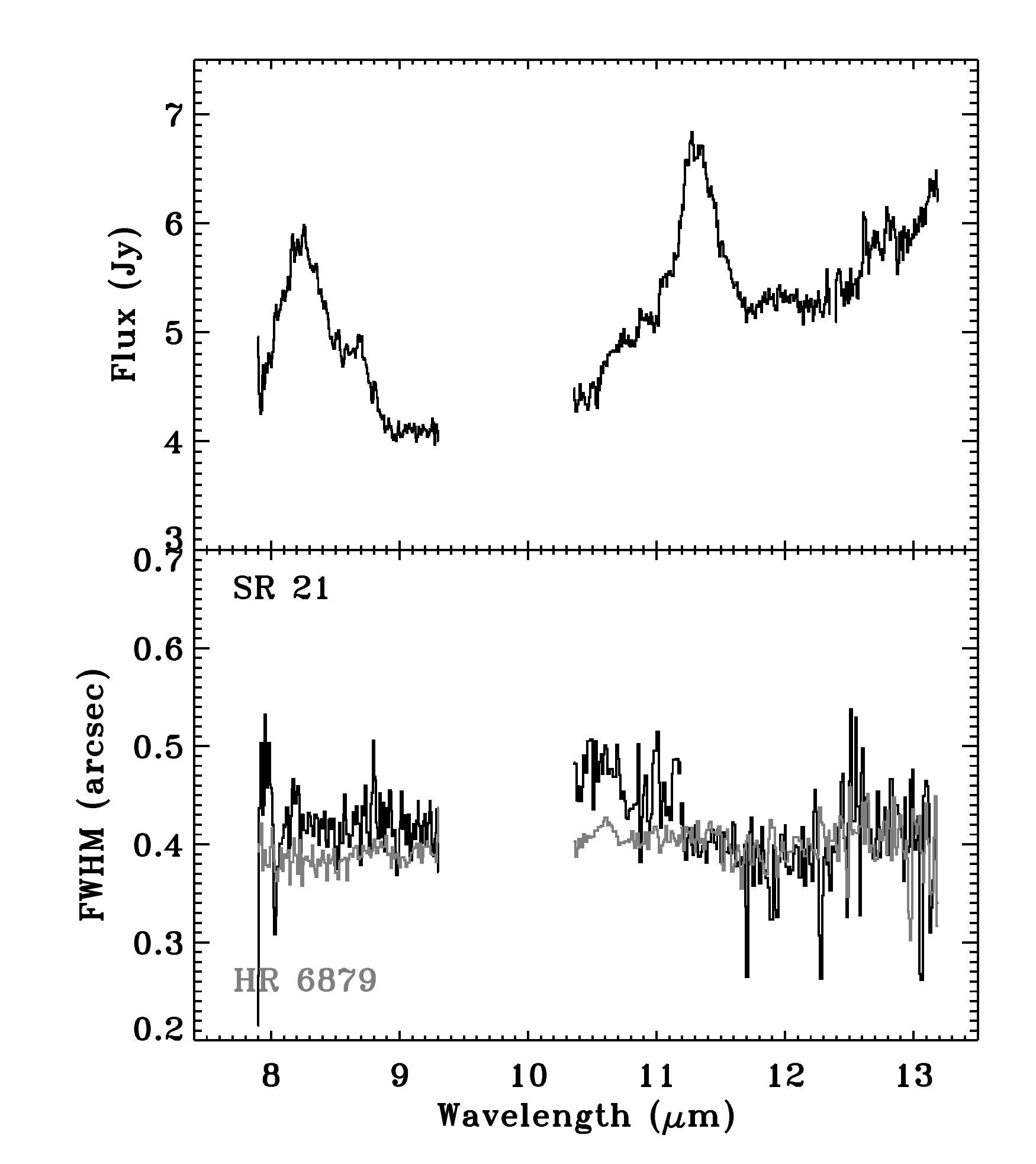}
  \caption{({\bf Upper panel}) VISIR N-band spectrum of SR~21A. ({\bf Lower panel}) FWHM of spatial profile of SR~21A and standard star HR~6879.}
  \label{fig:sr21visir}
\end{figure}
In contrast with WL~16, SR~21A is a source with very strong 3.3, 8.6, and 11.2 \micron\ features but spatially unresolved. The 3.3, 8.6--11.2 \micron\ disk continuum emission has a radial extent of $\leq$ 0.15-0.33$''$, corresponding to $\sim$19--41 AU, shown in Figs.~\ref{fig:sr21} and \ref{fig:sr21visir}, and in the left panel of Fig.~\ref{fig:wl16_2dspec}. This source has been interpreted as a `cold disk' source by \citet{bro07} with a gap between 0.45 and 18 AU, but with a large outer disk. The origin of the PAH emission is suspected to arise at the outer edge of the inner disk or within the gas filling the gap, inside 30 AU radius. The unknown position angle may have led to a chance alignment of the slit close to the semi-minor axis of the disk, which may have reduced the apparent spatial extent. 

\subsubsection{IRS~48}
\begin{figure}
  \centering
  \includegraphics[width=\columnwidth]{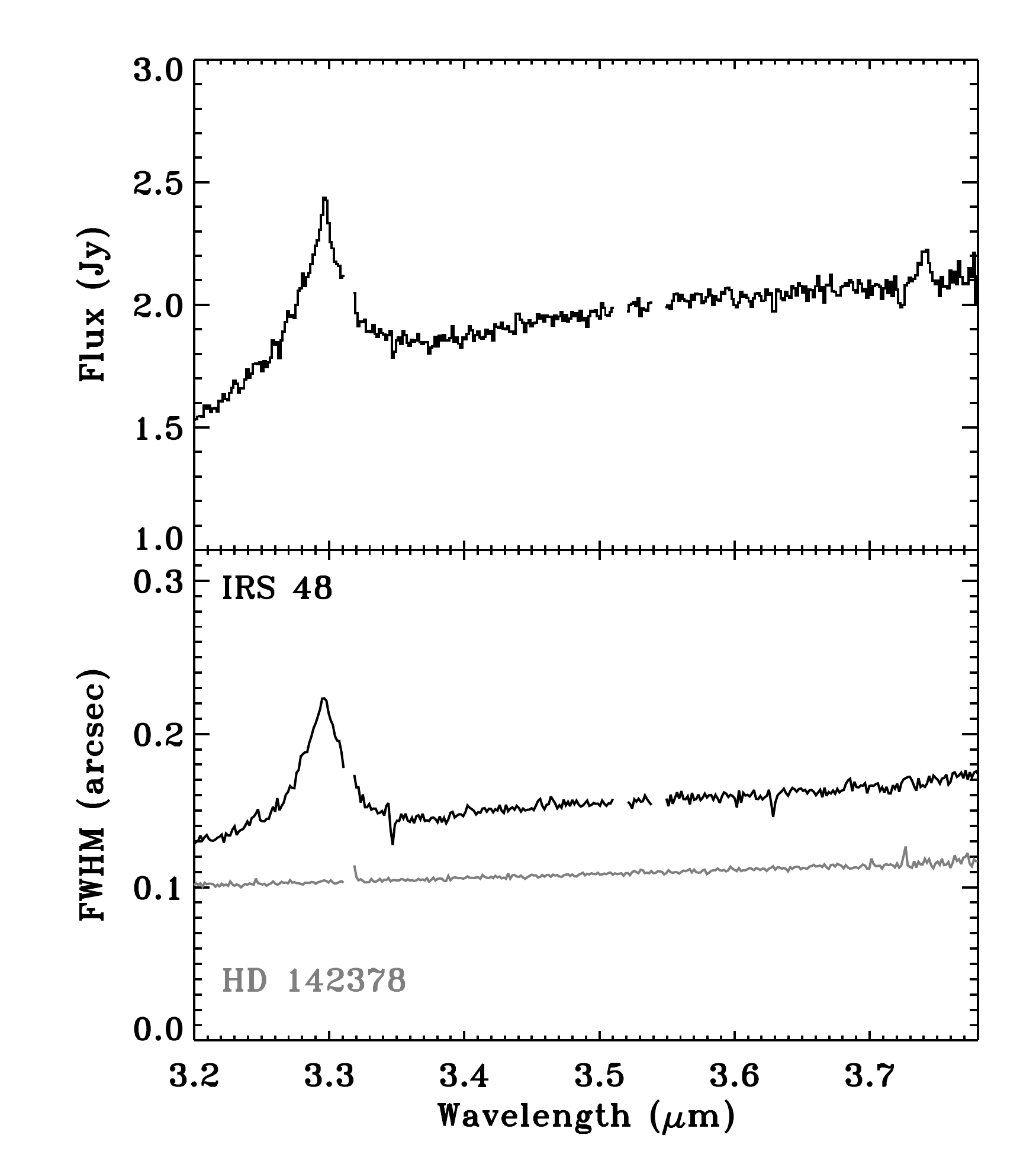}
  \caption{({\bf Upper panel}) NACO L-band spectrum of IRS~48. ({\bf Lower panel}) FWHM of spatial profile of IRS~48 and standard star HD~142378.}
  \label{fig:irs48naco1}
\end{figure}
IRS~48 is the only source in this sample for which all PAH features are spatially resolved. IRS~48 is a low-mass M0 star in the $\rho$ Oph cloud for which strong 3.3, 8.6, 11.2, 11.9, and 12.7 \micron\ PAH features are detected. For this source VISIR images have shown the presence of a 30 AU radius gap in the inner disk \citep{gee07} with the PAH feature emission originating from a region inside the gap. The disk continuum emission at 3--4 \micron\ in the NACO spectrum is spatially resolved with a radial extent rising from 0.06--0.09$''$ (7.5--11 AU) with wavelength. The 3.3 \micron\ PAH feature is detected and spatially resolved on a radial scale of 0.11$''$ (or 15 AU) above the extent of the continuum of $\sim 0.07''$ or 9 AU (Figs.\ \ref{fig:spatprof} and \ref{fig:irs48naco1}). Assuming diffraction limit and no influence of seeing and correcting for the contribution of the PSF based on the spatial extent of the underlying continuum, the radial extent of the 3.3 \micron\ features is found to be $\sim$ 11 AU. 
The disk continuum emission at 8--13 \micron\ in the VISIR spectrum is resolved with a radial extent increasing from 0.25$''$--0.33$''$ (31--41 AU). The 8.6, 11.2, and 12.7 \micron\ features are spatially resolved on a radial scale {\it smaller} than the surrounding disk continuum, with radial extents of 0.25, 0.32, and 0.33$''$ (31, 40, 41 AU) respectively, half-width at half-maximum (HWHM). These extents are consistent with the full (99\%) radial 8.6 and 11.2 \micron\ PAH extent of 75 and 90 AU respectively, found in spatially resolved in VISIR images in \citet{gee07}. Also, our finding of the smaller feature extent compared with the continuum is consistent with the VISIR images showing PAH emission to be smaller than the mid-IR 18.7 \micron\ emission.

\subsubsection{T~Cha}
For T~Cha, \citet{anc98} measured a distance of only 66 pc, making it one of the closest, isolated T Tauri stars, and not part of the Chamaeleon cloud ($d =178$ pc). A weak 11.2 \micron\ feature and a marginal 3.3 \micron\ feature are detected. The 3.3 \micron\ feature is unresolved, with a radial spatial extent of the continuum emission of $\leq$0.34$''$, corresponding to $\leq$ 22 AU. The 11.2 \micron\ feature is also unresolved, with a continuum spatial extent of $\leq$0.19$''$, corresponding to a radial extent of $\leq$13 AU.

\subsubsection{Summary and discussion}Ê
Table \ref{tab:featextent} summarizes our detections of spatially resolved PAH features. It tabulates the half-width at half-maximum extent of the emission, assuming it can be fitted with a Gaussian profile. Under this assumption, this HWHM extent corresponds to the radius of the disk, inwards of which 76\% of the normalized cumulative emission originates.

\begin{figure}
  \centering
  \includegraphics[width=\columnwidth]{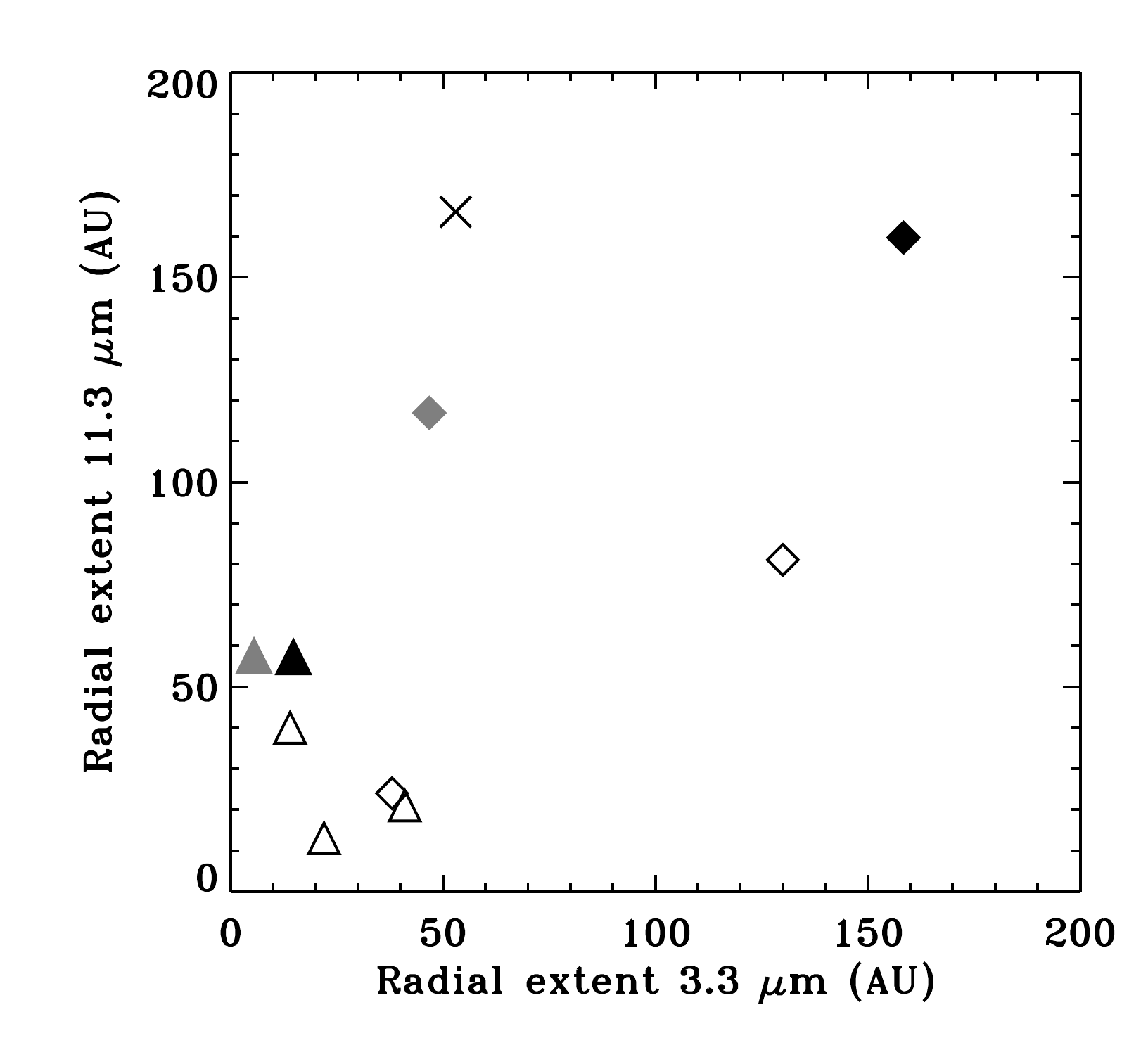}
  \caption{Radial extent of 3.3 \micron\ feature versus 11.2 \micron\ feature; Herbig Ae stars indicated by filled diamonds, T Tauri stars by filled triangles; models with $N_{\mathrm{c}}$ = 96 indicated by grey, $N_{\mathrm{c}}$=50 by black. Observations of IRS~48, SR~21A, T~Cha, HD~101412 and HD~98922 are indicated by open symbols. Model prediction point extracted from Fig.~4 of \citet{hab04} for a disk around a Herbig star for $N_{\mathrm{c}}$ = 100 is indicated by a cross.}
  \label{fig:modobs}
\end{figure}

This HWHM extent is compared with model predictions by \citet{vis07}, reproduced in Fig.\ \ref{fig:modelspex} in Appendix \ref{sec:appa}, with an improved extraction technique. Figure\ \ref{fig:modobs} plots the radial extent of the 3.3 \micron\ versus the 11.2 \micron\ feature, for both model predictions of Herbig Ae stars and T Tauri stars, for both small ($N_{\mathrm{c}}$=50) and large ($N_{\mathrm{c}}$=96) PAHs. 
The model extent of the 3.3 \micron\ feature varies strongly with PAH size, decreasing by a factor 2-3 when the PAH size is increased by a factor of $\sim 2$, while the extent of the 11.2 \micron\ feature is less sensitive.
For Herbig Ae stars, the model predicts for small PAHs ($N_{\mathrm{c}}$=50) that $\sim$76\% of the PAH emission should originate from a ring with a typical radius of $\sim$160 AU for 3.3 and 11.2 \micron\ and 175--190 AU for 6.2, 7.7, and 8.6 \micron. 
For larger PAHs ($N_{\mathrm{c}}$=96) the predicted distribution is more centrally peaked, with 76\% of the emission arising inwards of radii of 47, 96, 105, 103, and 117 AU for the 3.3, 6.2, 7.7, 8.6, and 11.2 \micron\ features respectively. 
In T Tauri disks the small PAH emission is also more centrally peaked, within radii of 15 AU for the 3.3 \micron\ feature, and 50--60 AU for the 6.2, 7.7, 8.6, and 11.2 \micron\ features.

The observed extents for IRS~48, SR~21A, T~Cha, HD~101412, and HD~98922 are included in Fig.\ \ref{fig:modobs}. Most PAH detections have upper limits on the radial spatial extent of typically 20-40 AU (with the exception of HD~98922, whose distance is uncertain) while for a few sources the PAH emission is spatially resolved with typical extent of 12--55 AU. One source, IRS~48, has spatially resolved detections of all features. 

For all observations with detected PAH features, the (upper limit for the) radial extent of the 11.2 \micron\ feature is a factor of 1.5--6 lower than derived from the prediction of the template model and we do not see any evidence for ring-shaped PAH emission as the model predicts. Within the \citet{vis07} models, this implies larger PAHs of typically 100 or more carbon atoms. 
The model prediction by \citet{hab04} for Herbig Ae stars with large PAHs ($N_{\mathrm{c}}$=100) is also indicated in Fig.\ \ref{fig:modobs}. Our measurement for the Herbig Ae star, HD~101412, is a factor 6--7 smaller than their and our prediction.

The typical extent of the 3.3 \micron\ feature at 76\% of 12--55 AU agrees with the predictions of both \citet{vis07} and \citet{hab04}.  Our derived upper limits on the extent of the 8.6 and 11.2 \micron\ features of typically $\leq$ 30 AU, are about a factor 2 smaller than the model predictions, which also hints at the presence of larger PAHs than considered in those models. As noted by \citet{gee06}, a significant number of these PAH sources are known to have substantial gaps out to 40 AU in their disks \citep{bro07}. The presence of warm 3 \micron\ disk emission suggests these sources have gaps, not holes. These observations put the PAH emission in the inner part of the disk, and in some cases (i.p.\ HD~100546) it very likely comes from the dust at the outer edge of the gap.

Four out of the five sources with spatially unresolved PAH features (leaving out HD~98922) have average upper limits on the extent of 30--40 AU, with no apparent difference between the Herbig Ae star HD~101412 and the 3 T Tauri stars. Potential misalignment of the slit with the semi-major axis of an inclined disk will reduce the measurable spatial extent (see also Appendix \ref{sec:appa}). In addition, unresolved features could have at least two geometrical causes. A larger inclination of the disk (from face-on) would decrease the measurable extent, although at near-IR wavelengths, for close to edge-on disks, scattering could dominate the spatial extent of the source of emission. Four out of the five unresolved sources (excluding EC~82) have no indications of a strong inclination in their SED \citep{kes06}. 
The second geometrical effect is the flaring or the covering fraction of the disk. For a disk with a decreasing covering fraction (through decreasing scale height), the feature to continuum ratio of the PAH features will decrease for very flat disks \citep{gee06}. Both HD~101412 and HD~98922 have SEDs consistent with a flat type II source \citep{mee01}, while the SR~21A and T~Cha both have SEDs consistent with still moderately large scale height, although the SED interpretation is more complicated due to the suspected presence of  a gap \citep{bro07}. 
Besides geometrical and instrumental effects, the alternative physical interpretation for the small extent could be the presence of predominantly large PAHs, which model predictions show to radiate mostly from smaller scales, as discussed above \citep{vis07}.

\section{Conclusions}
We detect the 3.3, 8.6, and 11.2 \micron\ PAH features in a small fraction of our sample of T Tauri stars, with typical upper limits between 1 $\times 10^{-15}$ and $5 \times 10^{-17}$ W m$^{-2}$. Compared with the Spitzer survey in \citet{gee06}, we confirm all their 11.2 \micron\ detections and find marginal-to-clear 3.3 \micron\ detections. The bias in our sample prevents us from deriving an independent detection fraction, but the low number of PAH observations toward T Tauri stars is consistent with the low detection rate of 8\% by \citet{gee06}. In two specific sources, WL~16 and HD~141569, the PAHs are expected to be largely ionized because of the strong 7.7 and 8.6 \micron\ features and lack of 3.3 \micron. 

The spatial extent of the PAH features is shown to be confined to scales smaller than $0.1-0.4''$ (HWHM), corresponding to radial scales of 12--60 AU in the disk, at typical distances of 150 pc, barring exceptional cases with unknown or much larger distances. In a few examples, the PAH features are resolved to be more extended than the hot underlying dust continuum of the disk, whereas in one case, IRS~48, the extent of the PAH emission is confirmed to be less than that of the large grains. The typical extent of the PAH features of 15--60 AU is found to be very similar for both Herbig Ae and T Tauri stars, and similar for all features. For Herbig Ae stars, the small 12--55 AU extent and absence of any ring emission is consistent with the model predictions of larger ($\geq$100 carbon atoms) grains. The same conclusion of a need for large PAHs holds for the T Tauri stars where the 8.6 and 11.2 \micron\ features appear to be smaller than predicted, although here the 3.3 \micron\ extent is consistent with smaller ($\sim$ 50 carbon atoms) PAHs. Given the large fraction of disks with gaps and PAHs, future modeling studies of the PAH extent should include the presence of gaps in disks.

\begin{acknowledgements}
KMP is supported by NASA through Hubble Fellowship grant 01201.01 awarded by the STScI, which is operated by the AURA, for NASA, under contract NAS 5-26555. Astrochemistry in Leiden is supported by a Spinoza grant from the Netherlands Organization for Scientific Research (NWO).
\end{acknowledgements}

\bibliographystyle{aa}
\bibliography{8466}

\appendix
\section{Spatial extent models}
\label{sec:appa}
\begin{figure}
\includegraphics[width=\columnwidth]{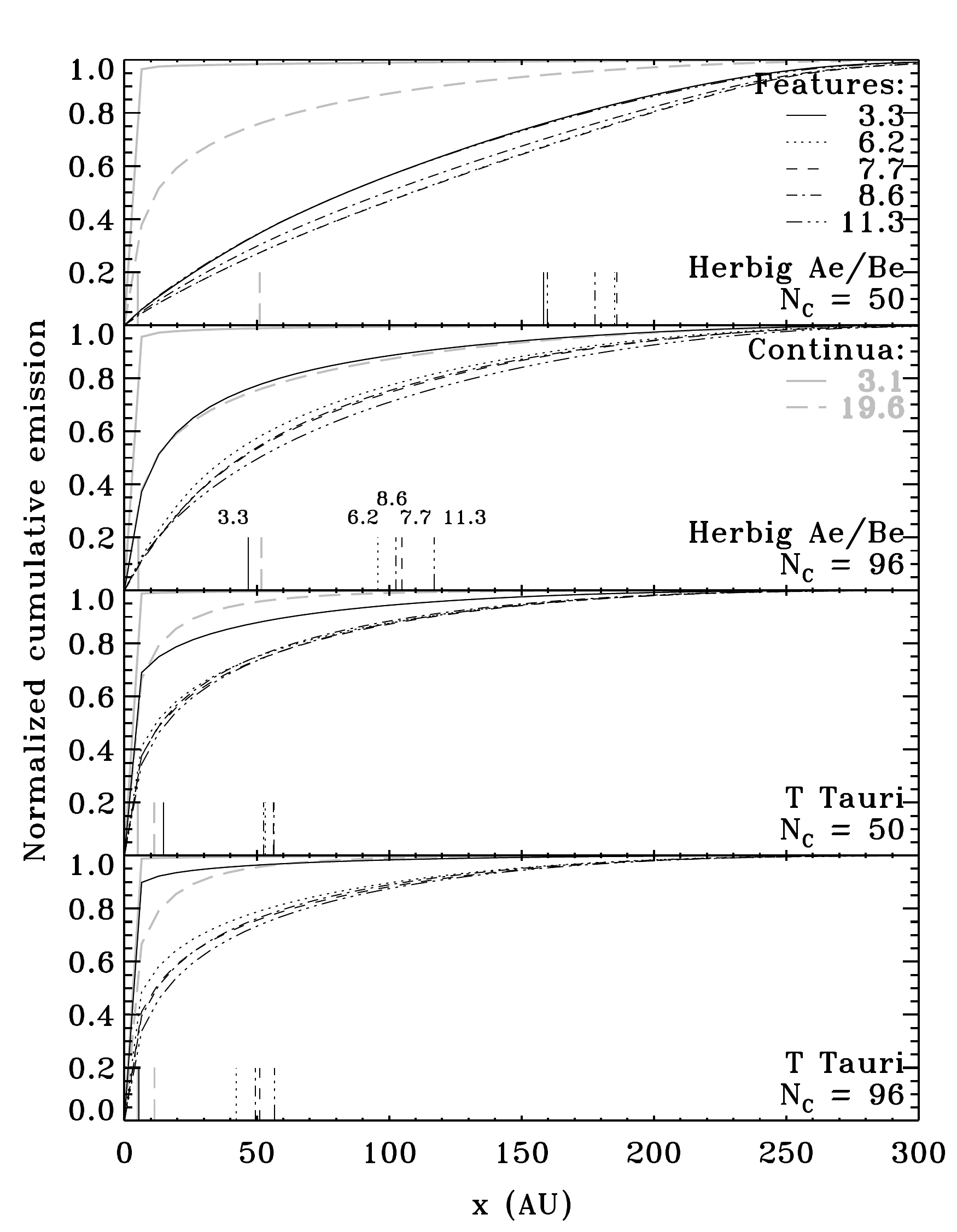}
\caption{Normalized cumulative intensity of the five main PAH features (black) and the continua at $3.1$ and $19.6$ \micron{} (gray) for $\ch{50}{18}{}$ and $\ch{96}{24}{}$ in the standard Herbig Ae/Be and T Tauri model disks (including PAH chemistry) from \citet{vis07}. The disks were synthetically observed face-on through an infinitely wide slit. The vertical bars indicate where each curve reaches $76\%$, corresponding to the HWHM of the Gaussians in Sect.\ \ref{ssec:measspatext}.}
\label{fig:modelspex}
\end{figure}

\citet[ hereafter V07]{vis07} modeled the chemistry of and infrared emission from PAHs in circumstellar disks, including an analysis of the spatial extent of the emission. 
The algorithm to determine the spatial extent of the emission was further improved to allow for a more direct comparison with observed spatial profiles by summing the emission in a (narrow) slit rather than in concentric rings as done in V07. A simplification was introduced by measuring continuum-subtracted peak intensities instead of spectrally integrated feature fluxes. Since the model features all have the same shape \citep{draine07a}, this produces the same spatial profiles. The new procedure is essentially the same as outlined in Sect.\ \ref{ssec:measspatext} for the observations. The improved profiles for the standard Herbig Ae/Be and T Tauri models ($R_{\mathrm{disk,in}}$ = 0.077 AU, $R_{\mathrm{disk,out}}$ = 300 AU, $M_{\mathrm{disk}}$ = 0.01 M$_{\odot}$) from V07, synthetically observed face-on through a slit wide enough to cover the entire disk, are presented in Fig.\ \ref{fig:modelspex}.

\begin{figure}
\includegraphics[width=\columnwidth]{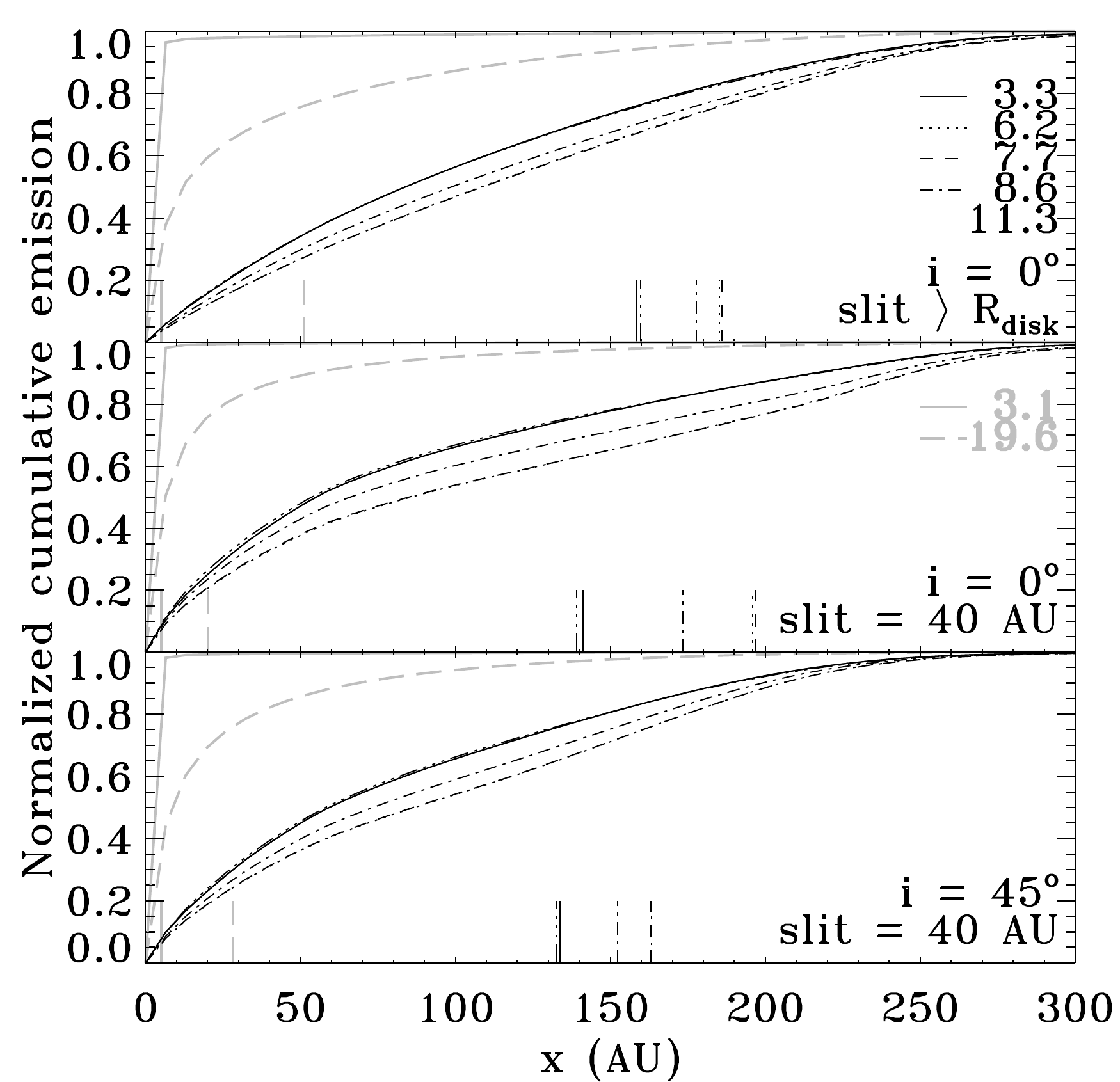}
\caption{Normalized cumulative intensity for $\ch{50}{18}{}$ in the standard Herbig Ae/Be model disk (including PAH chemistry) from \citet{vis07}. The disk was synthetically observed at $45^\circ$ or through a narrow slit as indicated. Lines and colors are as in Fig.\ \ref{fig:modelspex}.}
\label{fig:slitincl}
\end{figure}

If the source does not fit inside the width of the slit, part of the emission originating at large radii is blocked, thus generally reducing the apparent spatial extent compared to observations through a wider slit. This is shown in Fig.\ \ref{fig:slitincl}. With a $40$ AU slit ($0.27''$ at $150$ pc), the location inside of which $76\%$ of the feature emission originates shifts to smaller radii by the order of $10-15\%$ compared to an infinitely wide slit.

The effects of inclination are less straightforward. For objects at an inclination of about $45^\circ$, the observed spatial profiles differ only slightly from the face-on profiles if an infinitely wide slit is used. With a narrow slit, part of the emission is obscured asymmetrically across the width of the slit and the emission can appear to become both more or less extended, depending on the details of the source. This also occurs when the source is viewed more edge on, regardless of the width of the slit.

 \end{document}